\documentclass{llncs}
\usepackage{todonotes}
\usepackage{epsfig, amsfonts, amsmath, amssymb,fullpage}
\usepackage[colorlinks]{hyperref}
\usepackage{wrapfig,color}
\usepackage{tabularx}
\usepackage[ruled,vlined]{algorithm2e}

\usepackage{amsfonts}

\newcommand{\set}[1]{\left\{ #1\right\}}
\newcommand{\gilt}{:}
\newcommand{\sodass}{\,:\,}
\newcommand{\setGilt}[2]{\left\{ #1\sodass #2\right\}}

\newcommand{\realrange}[2]{\left[#1, #2\right]}

\newcommand{\unitrange}[2]{\realrange{0}{1}}

\newcommand{\llabel}[1]{\label{\labelprefix:#1}}
\newcommand{\labelprefix}{} 

\newcommand{\discussionsize}{\small}

\newcommand{\frage}[1]{}

\newenvironment{code}{\noindent
\begin{tabbing}%
\hspace{2em}\=\hspace{2em}\=\hspace{2em}\=\hspace{2em}\=\hspace{2em}\=%
\hspace{2em}\=\hspace{2em}\=\hspace{2em}\=\hspace{2em}\=\hspace{2em}\=%
\kill}{\end{tabbing}}

\newcommand{\labelcommand}{}
\newcommand{\captiontext}{}
\newsavebox{\codeparam}
\newcounter{lineNumber}
\newenvironment{disscodepos}[3]{%
\renewcommand{\labelcommand}{#2}%
\renewcommand{\captiontext}{#3}%
\sbox{\codeparam}{\parbox{\textwidth}{#3}}%
\begin{figure}[#1]\begin{center}\begin{code}\setcounter{lineNumber}{1}}{%
\end{code}\end{center}\caption{\llabel{\labelcommand}\captiontext}\end{figure}}

{\end{disscodepos}}

\newcommand{\Is}       {:=}

\newdimen\endofsize\endofsize=0.5em
\def\endofbeweis{~\quad\hglue\hsize minus\hsize
                 \hbox{\vrule height \endofsize width
\endofsize}\par}

\newcommand{\innerOuter}{\mathrm{innerOuter}}
\newcommand{\expansion}{\mathrm{expansion}}
\newcommand{\Outer}{\mathrm{Out}}

\newcommand{\nnzr}{\textsf{nnzr}}
\newcommand{\nnzc}{\textsf{nnzc}}

\def\MdR{\ensuremath{\mathbb{R}}}

\title{\Large Advanced Coarsening Schemes for Graph Partitioning\thanks{Partially supported by DFG SA 933/10-1 and CSCAPES institute, a DOE project.}}

\author{
Ilya Safro\inst{1}
\and Peter Sanders\inst{2}
\and Christian Schulz \inst{2}
}
\institute{ Mathematics and Computer Science Division, Argonne National Laboratory\\ \email{safro@mcs.anl.gov}
\and Karlsruhe Institute of Technology, Institute for Theoretical Informatics, Algorithmics II\\
\email{sanders@kit.edu,  christian.schulz@kit.edu} 
}
\date{}
\begin{document}

\maketitle
\begin{abstract}
The graph partitioning problem is widely used and studied in many practical and theoretical applications. 
The multilevel strategies represent today one of the most effective and efficient generic frameworks for solving this problem on large-scale graphs. 
Most of the attention in designing the multilevel partitioning frameworks has been on the refinement  phase. 
In this work we focus on the coarsening phase, which is responsible for creating structurally 
similar to the original but smaller graphs. 
We compare different matching- and AMG-based coarsening schemes, experiment with the algebraic distance between nodes, and demonstrate computational results on several classes of graphs that emphasize the running time and quality advantages of different coarsenings.
\end{abstract}
\section{Introduction}
\thispagestyle{empty}
Graph partitioning is a class of problems used in many fields of computer
science and engineering. Applications include VLSI design, load balancing for parallel computations,  network analysis, and optimal scheduling. The goal is to partition the vertices of a graph into a certain number of disjoint sets
of approximately the same size, so that a cut metric is minimized.
This problem is NP-complete even for several restricted classes of graphs, and there is no constant factor approximation algorithm for general graphs \cite{journals/ipl/BuiJ92}. 
In this paper we focus on a version of the problem that constrains the
maximum block size to $(1+\epsilon)$ times the average block size and tries to
minimize the total cut size, namely, the number of edges that run between blocks.

Because of the  practical importance, many heuristics of different nature (spectral \cite{pothen-part}, combinatorial \cite{fiduccia1982lth}, evolutionist \cite{buiMoon96,kaffpaE}, etc.) have
been developed to provide an approximate result in a reasonable (and, one hopes,
linear) computational time. We refer the reader to  \cite{fjallstrom1998agp,SchKarKum00,Walshaw07} for more material.
However, only the introduction of the general-purpose multilevel methods during the 1990s has provided a breakthrough in efficiency and quality. 
The basic idea can be traced back to multigrid solvers for solving elliptic partial differential equations \cite{mgbooktrott} but more recent practical methods are based on mostly graph-theoretic aspects of, in particular, edge contraction and local search.  Well-known software packages based on this approach include Jostle~\cite{Walshaw07}, Metis \cite{SchKarKum00}, DiBaP \cite{meyerhenke2008ndb}, and Scotch \cite{Scotch}.  

A multilevel algorithm consists of two main phases: \emph{coarsening} -- where the problem instance is  gradually mapped to smaller ones to reduce the original complexity (i.e., the graph underlying the problem is compressed), 
and \emph{uncoarsening} -- where the solution for the original instance is constructed by using the information inherited from the solutions created at the next coarser levels. 
 So far, most of the attention in designing the multilevel partitioning frameworks has been on the uncoarsening phase. In this work we focus on the coarsening phase, which is responsible for creating  graphs that are smaller than but structurally similar to the given graph. We compare different coarsening schemes, introduce new elements to them, and demonstrate computational results. For this purpose different coarsening schemes are integrated into the graph partitioning framework KaFFPa \cite{kaffpa}.

The paper is organized as follows. We begin in Section~\ref{sec:preliminaries} by introducing notation  and the multilevel approach. 
In Section~\ref{sec:coarsening} we describe different coarsening schemes, including a novel algebraic, multigrid-inspired balanced coarsening scheme and matching-based coarsening schemes, as well as new measures for connectivity. 
We present a large experimental evaluation in Section~\ref{sec:experiments} on graphs arising in real-world applications and on graphs that are specifically designed to be hard for multilevel algorithms.

\section{Preliminaries}
\label{sec:preliminaries}
Consider an undirected graph $G=(V,E,c,\omega)$
with edge weights\footnote{Subscripts will be used for a short notation; i.e., $\omega_{ij}$ corresponds to the weight of $\{i,j\}\in E$.} $\omega: E \to \MdR_{>0}$, node weights
$c: V \to \MdR_{\geq 0}$, $n = |V|$, and $m = |E|$.
We extend $c$ and $\omega$ to sets; in other words,
$c(V')\Is \sum_{v\in V'}c(v)$ and $\omega(E')\Is \sum_{e\in E'}\omega(e)$.
Here, $\Gamma(v)\Is \setGilt{u}{\set{v,u}\in E}$ denotes the neighbors of $v$.
We are looking for \emph{blocks} of nodes $V_1$,\ldots,$V_k$
that partition $V$, namely, $V_1\cup\cdots\cup V_k=V$ and $V_i\cap V_j=\emptyset$
for $i\neq j$. The \emph{balancing constraint} demands that
$\forall i\in \{1..k\}\gilt c(V_i)\leq L_{\max}\Is (1+\epsilon)c(V)/k+\max_{v\in V} c(v)$ for
some parameter $\epsilon$.
The last term in this equation arises because each node is atomic and
therefore a deviation of the heaviest node has to be allowed.
The objective is to minimize the total \emph{cut} $\sum_{i<j}\omega (E_{ij})$ where
$E_{ij}\Is\setGilt{\set{u,v}\in E}{u\in V_i,v\in V_j}$.
A vertex $v \in V_i$ that has a neighbor $w \in V_j, i\neq j$, is a boundary vertex.
We denote by $\nnzr(A,i)$ and $\nnzc(A,i)$ the number of nonzero entries in the $i$th row or column of a matrix $A$, respectively.

A matching $M\subseteq E$ is a set of edges that do not share any common nodes; that is, the graph $(V,M)$ has maximum degree one.  \emph{Contracting} an edge $\set{u,v}$ means replacing the nodes $u$ and $v$ by a new node $x$ connected
to the former neighbors of $u$ and $v$.
We set $c(x)=c(u)+c(v)$ so the weight of a node at each level is the number of nodes it is representing in the original graph. If replacing edges of the form $\set{u,w}$,$\set{v,w}$ would generate two parallel edges $\set{x,w}$, we insert a single edge with
$\omega(\set{x,w})=\omega(\set{u,w})+\omega(\set{v,w})$.
\emph{Uncontracting} an edge $e$ undoes its contraction.

\subsection{Multilevel Graph Partitioning}
\begin{wrapfigure}{l}{7cm}
\centering
\vspace*{-.5cm}
\includegraphics[width=7cm]{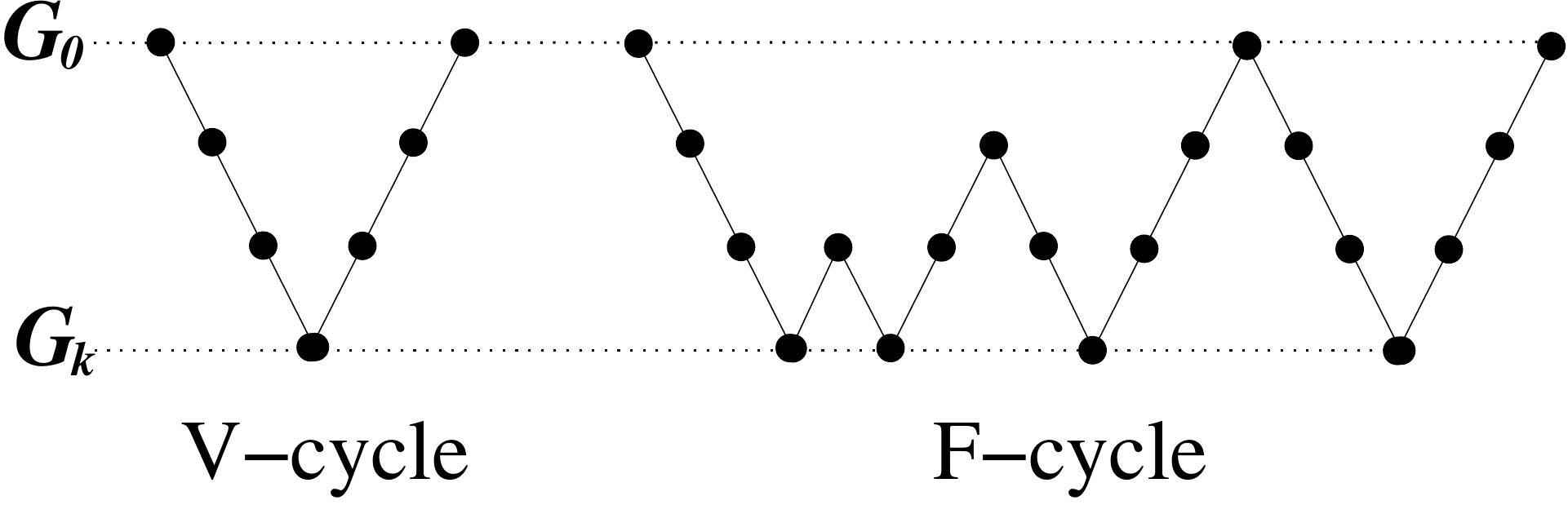}
\caption{V- and F-cycles schemes.}\label{fig:cycles}
\end{wrapfigure}

In the multilevel framework we construct a hierarchy of decreasing-size graphs, $G_0,~G_1, \dots , G_k$,  by \emph{coarsening}, starting from the given graph $G_0$ such that each next-coarser graph $G_i$ reflects basic properties of the previous graph $G_{i-1}$. 
At the coarsest level $G_k$ is partitioned by a hybrid of external solvers, and starting from the $(k-1)$th level the solution is projected gradually (level by level) to the finest level. 
Each projection is followed by the \emph{refinement}, which moves nodes between the blocks in order to reduce the size of the cut.  
This entire process is called a V-cycle (see Figure \ref{fig:cycles}). KaFFPa \cite{kaffpa} extended the
concept of \emph{iterated multilevel algorithms}  which 
 was introduced for graph partitioning by Walshaw et al. \cite{walshaw2004multilevel}. 
The main idea is to iterate the multilevel-scheme using different random seeds for coarsening and uncoarsening. This ensures non-decreased partition quality since the refinement algorithms of KaFFPa guarantee no worsening.
In this paper, for the purpose of comparison we consider also F-cycles \cite{kaffpa} (see Figure \ref{fig:cycles})  as a potentially stronger and slower version of the multilevel framework for the graph partitioning problem. 
The detailed description of F-cycles for the multilevel graph partitioning framework can be found in \cite{kaffpa}.

\section{Coarsening Schemes}
\label{sec:coarsening}

\par One of the most important concerns of multilevel schemes is a measure of the connection strength between vertices. 
For matching-based coarsening schemes, experiments indicate that more sophisticated \textit{edge rating} functions are superior to edge weight as a criterion for the matching algorithm \cite{kappa}. 
To be more precise first the edges get rated using a rating function that indicates how much sense it makes to contract an edge. Then a matching algorithm tries to maximize the sum of the ratings of the edges to be contracted. 
The default configurations of KaFFPa employ the ratings 
\begin{eqnarray*}
\expansion^{*2}(\set{u,v}) & \Is &  {\omega(\set{u,v})^2}/{c(u)c(v)}, \text{ and}\\
\innerOuter(\set{u,v}) & \Is & {\omega(\set{u,v})}/{(\Outer(v)+\Outer(u)-2\omega({u,v}))},
\end{eqnarray*}
where $\Outer(v)\Is \sum_{x\in\Gamma(v)}\omega(\set{v,x})$, since they yielded the best results in \cite{kappa}. 
\paragraph*{Algebraic distance for graph partitioning.} The notion of algebraic distance introduced in \cite{safro:relaxml,chen-safro-algdist-full} is based on the principle of obtaining low-residual error components used in the Bootstrap AMG \cite{bamg-review}. 
When a priori knowledge of the nature of this error is not available, slightly relaxed random vectors are used to approximate it. This principle was used for linear ordering problems to distinguish between local and global edges \cite{safro:relaxml}. The main difference between the $k$-partitioning problem and other (not necessarily combinatorial) problems for which the algebraic distance has been tested so far is the balancing constraints. For many instances, it is important to keep the coarsening balanced; otherwise, even though the structural information will be captured by a sophisticated coarsening procedure, most of the actual computational work that constructs the approximate solution will be done by the refinement iterations. Bounding the number of refinement iterations may dramatically decrease its quality. 
Thus, a volume-normalized algebraic distance is introduced 
to take into account the balancing of vertices. 
\par Given the Laplacian of a graph $L=D-W$, where $W$ is a weighted adjacency matrix of a graph and $D$ is the diagonal matrix with entries $D_{ii}=\sum_j\omega_{ij}$, we define its volume-normalized version denoted by $\tilde{L}=\tilde{D}-\tilde{W}$ based on volume-normalized edge weights $\tilde{\omega}_{ij} = \omega_{ij}/\sqrt{c(i)c(j)}$. We define an iteration matrix $H$ for Jacobi over-relaxation (also known as a lazy random-walk matrix) as
\[
 H = (1-\alpha)I + \alpha \tilde{D}^{-1}\tilde{W},
\]
where $0\leq \alpha \leq 1$. The algebraic distance coupling $\rho_{ij}$ is defined as
\[
\rho_{ij} = \big( \sum_{r=1}^R |\chi_i^{(k,r)} - \chi_j^{(k,r)}|^2 \big)^\frac{1}{2}~,
\]
where $\chi^{(k,r)} = H^k \chi^{(0,r)}$ is a relaxed randomly initialized test vector (i.e., $\chi^{(0,r)}$ is a random vector sampled over \text{[-1/2, 1/2]}), $R$ is the number of test vectors, and $k$ is the number of iterations. In our experimental settings we set $\alpha=0.5$, $R=5$, and $k=20$.

\subsection{Coarsening}
\par To the best of our knowledge, the existing multilevel algorithms for combinatorial optimization problems (such as $k$-partitioning, linear ordering, clustering, and segmentation) can be divided into two classes: contraction-based schemes \cite{kaffpa,Dhillon05afast,KarypisK95} (including contractions of small subsets of nodes \cite{BartelGKM10}) and algebraic multigrid (AMG)-inspired schemes \cite{safro:relaxml,Hu:wavefront,sharon,doritpart}.

\subsubsection{AMG-inspired coarsening.}
\par One of the most traditional approaches for derivation of the coarse systems in AMG is the Galerkin operator \cite{mgbooktrott}, which projects the fine system of equations to the coarser scale. In the context of graphs this projection is defined as
\begin{equation}
L_c = P L_f P^T,
\end{equation}
where $L_f$ and $L_c$ are the Laplacians of fine and coarse graphs $G_f=(V_f,E_f)$ and $G_c=(V_c,E_c)$, respectively. The $(i,J)$th entry of projection matrix $P$ represents the strength of the connection between fine node $i$ and coarse node $J$. The entries of P, called interpolation weights, describe both the coarse-to-fine and fine-to-coarse relations between nodes. 
\par The coarsening begins by selecting a dominating set of (seed or coarse) nodes $C \subset V_f$ such that all other (fine) nodes in $F=V_f\setminus C$ are strongly coupled to $C$. This selection can be done by traversing all nodes and leaving node $i$ in $F$ (initially $F=V_f$, and $C=\emptyset$) that satisfy
\begin{equation}
\sum_{j\in C} 1/\rho_{ij}  \geq \Theta \cdot \sum_{j\in V_f} 1/\rho_{ij},
\end{equation}
where $\Theta$ is a parameter of coupling strength. As in AMG-based approaches for linear ordering problems \cite{safro2005} we observed that the order in which $V_f$ is traversed does play an important role in reducing the dependence on random seeds (for details on future volume ordering see \cite{safro:relaxml}).
\par The Galerkin operator construction differs from other AMG-based approaches for combinatorial optimization problems. Balancing constraints of the partitioning problem require a limited number of fine-to-coarse attractions between $i\in C$ ($i$th column in $P$) and its neighbors from $F$ (nonzero entries in the $i$th column in $P$). In particular, this is important for graphs where the number of high-degree nodes in $C$ is smaller than the number of parts in the desired partition. Another well-known problem of AMG that can affect the performance of the solver is the complexity of coarse levels. Consideration of the algebraic distance makes it possible to minimize the order of interpolation (the number of fractions a node from $F$ can be divided to) to 1 or 2 only \cite{safro:relaxml}. Algorithm \ref{alg:iop} summarizes the construction of $P$.

\vspace*{-.25cm}
\incmargin{1em}
\restylealgo{boxed}\linesnumbered
\begin{algorithm}
\SetKwData{Left}{left}
\SetKwFunction{Union}{Union}
\SetKwInOut{Input}{input}
\SetKwInOut{Output}{output}
\caption{Interpolation weights for $P$}\label{alg:iop}
\Input{$G$, $i\in V_f$, $P$}
\BlankLine
\If{$i\in C$}{
$P_{iI(j)} \leftarrow 1$\;
}
\Else{
$l \leftarrow $ list of at most $\kappa$ algebraically strongest connections of $i$ to $C$\;
$\{e_1,e_2\} \leftarrow $ algebraically strongest pair of edges (according to $\rho_{e_1}+\rho_{e_2}$) in $l$ such that the corresponding $C$-neighbors are not over-loaded if $i$ is divided between them\;
\If{$\{e_1,e_2\} \neq \emptyset$}{$l \leftarrow \{e_1,e_2\}$}
\Else{
$e_1 \leftarrow $ algebraically strongest connection of $i$ to $C$ such that the corresponding $C$-neighbor is not over-loaded if $i$ is aggregated with it\;
$l \leftarrow \{e_1\}$\;
}
\If{$l$ is empty}{move $i$ to $C$}
\Else{$N^c_i \leftarrow $ $C$-neighbors of $i$ that adjacent to edges in $l$\;
$P_{iI(j)} \leftarrow 1/(\rho_{ij}\cdot\sum_{k\in N^c_i}1/\rho_{ik}) \text{ for } j\in N^c_i$\;
update future volumes of $j\in N^c_i$\;
}
}
\end{algorithm}
\vspace*{-.5cm}
\decmargin{1em}
Algorithm \ref{alg:iop} can be viewed as simplified version of bootstrap AMG \cite{bamg-review}  with the additional restriction on future volume of aggregates and adaptive interpolation order. $P_{iI(j)}$ thus represents the likelihood of $i$ belonging to the $I(j)$th aggregate. The edge connecting two coarse aggregates $p$ and $q$ is assigned with the weight $w_{pq}=\sum_{k\not= l}P_{kp}w_{kl}P_{lq}$. The volume of the $p$th coarse aggregate is $\sum_j c(j)P_{jp}$. 
We emphasize the property of adaptivity of $C$ (line 15 in Algorithm \ref{alg:iop}), which is updated if the balancing of aggregates is not satisfied.
\par We mention the difference between our AMG scheme and the weighted aggregation (WAG) scheme in \cite{ChevalierS09}. The common principle that works in both schemes is based on the division of $F$-nodes between their $C$-neighbors.
 However, two critical components are missing in  \cite{ChevalierS09}: (1) the algebraic distance that forms both the set of seeds and the interpolation operator; and (2) the weight-balancing algorithmic component when aggregates  are created, namely, operator $P$ in \cite{ChevalierS09} is created as in classical AMG schemes. 
 One important disadvantage of \cite{ChevalierS09} is a relatively high  density of coarse levels, which is eliminated with introduction of the algebraic distance.
  This was achieved by reducing the order of interpolation to 1 or 2. 
  The balancing factor played an important role in reducing the running time of the algorithm. 
  Recently introduced max-flow/min-cut refinement leads to noticeably better results than FM/KL heuristics (explained in Section \ref{sec:uncoarsening}). In contrast to simple FM/KL swaps, however, its complexity becomes higher if the aggregates are unbalanced with respect to the maximum size of one part. Applying this refinement with unbalanced WAG can significantly increase the total running time of the solver or lead to weak solutions if the refinement is terminated before it finds a good local minimum. Overall, the performance of our AMG scheme differs significantly from what we observed with WAG.

\subsubsection{Matching based coarsening.}
\par Another coarsening framework, which is more popular because of its simplicity and faster performance, is the \emph{matching based scheme}. 
 In this scheme a coarse graph is constructed by using contractions derived from a preprocessed edge matching. This scheme represents a special case of $PL_fP^T$ in which $\nnzr(P,r)=1$ for all rows $r$ in $P$, and $1\leq \nnzc(P,c)\leq 2$ for all columns $c$ in $P$.
\paragraph{Global Paths Algorithm.}
The Global Paths Algorithm (GPA), was proposed in \cite{MauSan07} as a synthesis of Greedy and Path Growing algorithms \cite{DH03a}. 
Similar to the Greedy approach, GPA scans the edges in order of decreasing weight (or rating); 
but rather than immediately building a matching, it first constructs a collection
of paths and even length cycles. To be more precise, these paths initially contain no edges. 
While scanning the edges, 
the set is then extended by successively adding applicable edges. An edge is called applicable if it connects two endpoints of different paths or the two endpoints of an odd length path. Afterwards, optimal solutions are computed for each of these paths and cycles using dynamic programming. 
KaFFPaStrong \cite{kaffpa} employs $\innerOuter$ on the first level of the hierarchy since $\expansion^{*2}$ evaluates to one on unweighted graphs. Afterwards it uses $\expansion^{*2}$. 
\paragraph{RandomGPA Algorithm.}
This algorithm is used by the classic KaFFPaEco configuration. 
It is a synthesis of the most simple random matching algorithm and the GPA algorithm. 
To be more precise this matching algorithm is dependent of the number of blocks the graph has to be partitioned in. 
It matches the first $\max\{2, 7-\log k\}$ levels using the random matching algorithm and switches to the GPA algorithm afterwards. 
The random matching algorithm traverses the nodes in a random order and if the current node is not already matched it chooses a random unmatched neighbor for the matching.
KaFFPaEco employs $\expansion^{*2}$ as a rating function as soon as it uses GPA.
\subsection{The Coarsest Level}
Contraction is stopped when the graph is small enough to be partitioned by some other expensive algorithm. We use the same initial partitioning scheme as in KaFFPa \cite{kaffpa}, namely, the libraries Scotch and Metis for initial partitioning.
For AMG, some modifications have to be made since Scotch and Metis cannot deal with fractional numbers and Metis expects $\omega_{ij}\geq 1$. 
To overcome this implementational problem, we perform the following two steps. 
First, we divide each edge weight of the coarsest graph by the smallest edge weight that occurred on that level. 
This step assures edge weights larger than or equal to one without skewing the graph partitioning problem for the library used. 
Second, we get rid of the fractional edge weights using randomized rounding. Let $e \in E$ be an edge with fractional edge weight. We then obtain an integer edge weight $\tilde \omega (e)$  by flipping a coin with probabilities $\mathcal{P}(\text{head})= \omega(e)- \lfloor \omega(e) \rfloor,  \mathcal{P}(\text{tail}) = 1 - \mathcal{P}(\text{head})$. 
In the case of heads we set the edge weight $\tilde \omega(e)$ to $\lceil \omega(e) \rceil$; otherwise we set it to $\lfloor \omega(e) \rfloor$. This way we can assure that the value of the cut in the graph $\tilde G = (V_k, E_k, \tilde \omega)$ produced by the external initial partitioning algorithm is close to the real cut value in $G$. 

\subsection{Uncoarsening}\label{sec:uncoarsening}
Recall that uncoarsening undoes contraction. For AMG-based coarsening this means that fine nodes have to be assigned to blocks of the partition of the finer graph in the hierarchy. We assign a fine node $v$ to the block that minimizes cut$_B\cdot p_B(v)$, where cut$_B$ is the cut after $v$ would be assigned to block $B$ and $p_B(v)$ is a penalty function to avoid blocks that are heavily overloaded. To be more precise, after some experiments we fixed the penalty function to $p_B(v) = 2^{\max(0,100 \frac{c(B)+c(v)}{L_\text{max}})}$, where $L_\text{max}$ is the upper bound for the block weight.
Note that slight imbalances (e.g. overloaded blocks), can usually be fixed by the refinement algorithms implemented within KaFFPa.
For matching-based coarsening the uncoarsening is straightforward: 
a vertex is assigned to the block of the corresponding coarse vertex.

\paragraph*{Karlsruhe Fast Flow Partitioner (KaFFPa).}
Since we integrated different coarsening schemes into the multilevel graph partitioner KaFFPa \cite{kaffpa}, we now briefly outline the techniques KaFFPa uses during uncoarsening. 
After a matching is uncontracted, local search-based refinement algorithms move nodes between block boundaries in order to reduce the cut while maintaining the balancing constraint. 
Local improvement algorithms are usually variants of the FM algorithm \cite{fiduccia1982lth}. 
The variant KaFFPa uses is organized in rounds. In each round, a priority queue $P$ is used that is initialized with all vertices that are incident to more than one block, in a random order. 
The priority is based on the gain $g(i) = \max_P g_P(i)$ where $g_P(i)$ is the decrease in edge cut when moving $i$ to block $P$.  
Local search then repeatedly looks for the highest gain node $v$ and moves it to the corresponding block that maximizes the gain.
However, in each round a node is moved at most once. 
After a node is moved, its unmoved neighbors become eligible, i.e. its unmoved neighbors are inserted into the priority queue. 
When a stopping criterion is reached, all movements to the best-found cut that occurred within the balance constraint are undone.
This process is repeated several times until no improvement is found.

\paragraph*{Max-Flow Min-Cut Local Improvement.} During the uncoarsening phase KaFFPa additionally uses more advanced refinement algorithms. 
The first method is based on  max-flow min-cut computations between pairs of blocks, in other words, a method to improve a given bipartition. 
Roughly speaking, this improvement method is applied between all pairs of blocks that share a nonempty boundary. 
The algorithm basically constructs a flow problem by growing an area around the given boundary vertices of a pair of blocks such that each $s$-$t$ cut in this area yields a feasible bipartition of the original graph/pair of blocks \textit{within} the balance constraint. 
One can then apply a max-flow min-cut algorithm to obtain a min-cut in this area and therefore a nondecreased cut between the original pair of blocks. 
This can be improved in multiple ways, for example, by iteratively applying the method, searching in larger areas for feasible cuts, and applying most balanced minimum cut heuristics. For more details we refer the reader to \cite{kaffpa}.

\paragraph*{Multi-try FM.} 
The second method for improving a given partition is called multi-try FM. 
This local improvement method moves nodes between blocks in order to decrease the cut. 
Previous $k$-way methods were initialized with \textit{all} boundary nodes, i.e., all boundary nodes were eligible for movement at the beginning.
Roughly speaking, the multi-try FM algorithm is a $k$-way improvement method that is initialized with a \textit{single} boundary node, thus achieving a more localized search. This is repeated several rounds. 
For more details about the multi-try FM algorithm we refer the reader to \cite{kaffpa}. 

\section{Experimental Evaluation}
\label{sec:experiments}
\paragraph*{Configurations of KaFFPa.} 
The AMG coarsening was implemented separately based on the coarsening for linear ordering solvers from \cite{safro:relaxml} and was integrated into KaFFPa \cite{kaffpa}.
The computational experiments have been performed with six configurations of KaFFPa, which are presented in Table~\ref{tab:configurations}. 
All configurations use the described FM algorithm and flows for the refinement. The strong configurations further employ flows using larger areas, multi-try FM and F-cycles. A detailed description of the refinement configurations can be found in \cite{kaffpa}. Throughout this section, because of the respective similar running times, we concentrate on two groups of comparison: for fast versions (AMG-ECO, ECO, ECO-ALG) and for strong versions (AMG, STRONG, F-CYCLE). To be precise, usually
the running time of F-CYCLE is bigger than that of STRONG and AMG. However, the running time gap between fast and strong versions is even more significant on average. Since the main goal of this paper is to introduce the AMG coarsening with different uncoarsening configurations, most of the comparisons will be of type AMG vs respective non-AMG ratios. A comprehensive comparison of the F-CYCLE and the STRONG configuration can be found in \cite{kaffpa}. 

\par All experiments are performed with fixed imbalance factor 3\%. We also checked other small values, namely, 0\%, 1\%, and 5\%; however, no significant difference in the comparison of the respective methods was observed.
\vfill
     \begin{table}[h!]

        \small

        \vspace{-.25cm}
        \begin{tabularx}{\textwidth}{|l|X|}
        
        \hline
 ECO  &Represents the classical KaFFPaEco configuration, a good trade-off of partition quality and runtime. \\
         \hline
ECO-ALG &  Same refinement as in ECO, coarsening uses the GPA algorithm at each level and the edge rating function employs algebraic distances; i.e., it uses the rating function ex\_alg$(e) := \expansion^{*2}(e)/\rho_e$.  \\
         \hline
F-CYCLE & Represents the classical KaFFPaStrong configuration using strong refinement schemes and the F-cycle scheme, with the purpose of achieving high partition quality; this configuration achieved the best known partitions for many instances from Benchmark I in 2010 \cite{kaffpa}.\\
         \hline
STRONG & Uses the same refinement and matching schemes as in the F-CYCLE configuration; however, here only one single V-cycle is performed. \\
         \hline
AMG-ECO & AMG coarsening based on algebraic distances with interpolation order at most 2, refinement as in ECO. \\
         \hline
AMG & Same coarsening as in AMG-ECO, same refinement as in STRONG.\\
        \hline
        \end{tabularx}
        \vspace{.25cm}
        \caption{Description of the six configurations used for the computational experiments.}
        \label{tab:configurations}
        \vspace{-.75cm}
        \end{table}

\paragraph{Benchmark I: Walshaw's Partitioning Archive.} Chris Walshaw's benchmark archive \cite{soper2004combined} is a collection of real-world instances for the partitioning problem. The rules used there imply that the running time is not an issue, but one wants to achieve minimal cut values for $k \in \{2,4,8,16,32,64\}$ and balance parameters $\epsilon \in \{0,0.01,0.03,0.05\}$. It is the most used graph partitioning benchmark in the literature. 
Most of the graphs of the benchmark come from finite-element applications; however, there are also some graphs from VLSI design and a road network. Over the years many different heuristics have been tested and adjusted on this benchmark, so that many heuristics are able to obtain good results on these graphs.
\par
\begin{figure}
\includegraphics[width=8cm]{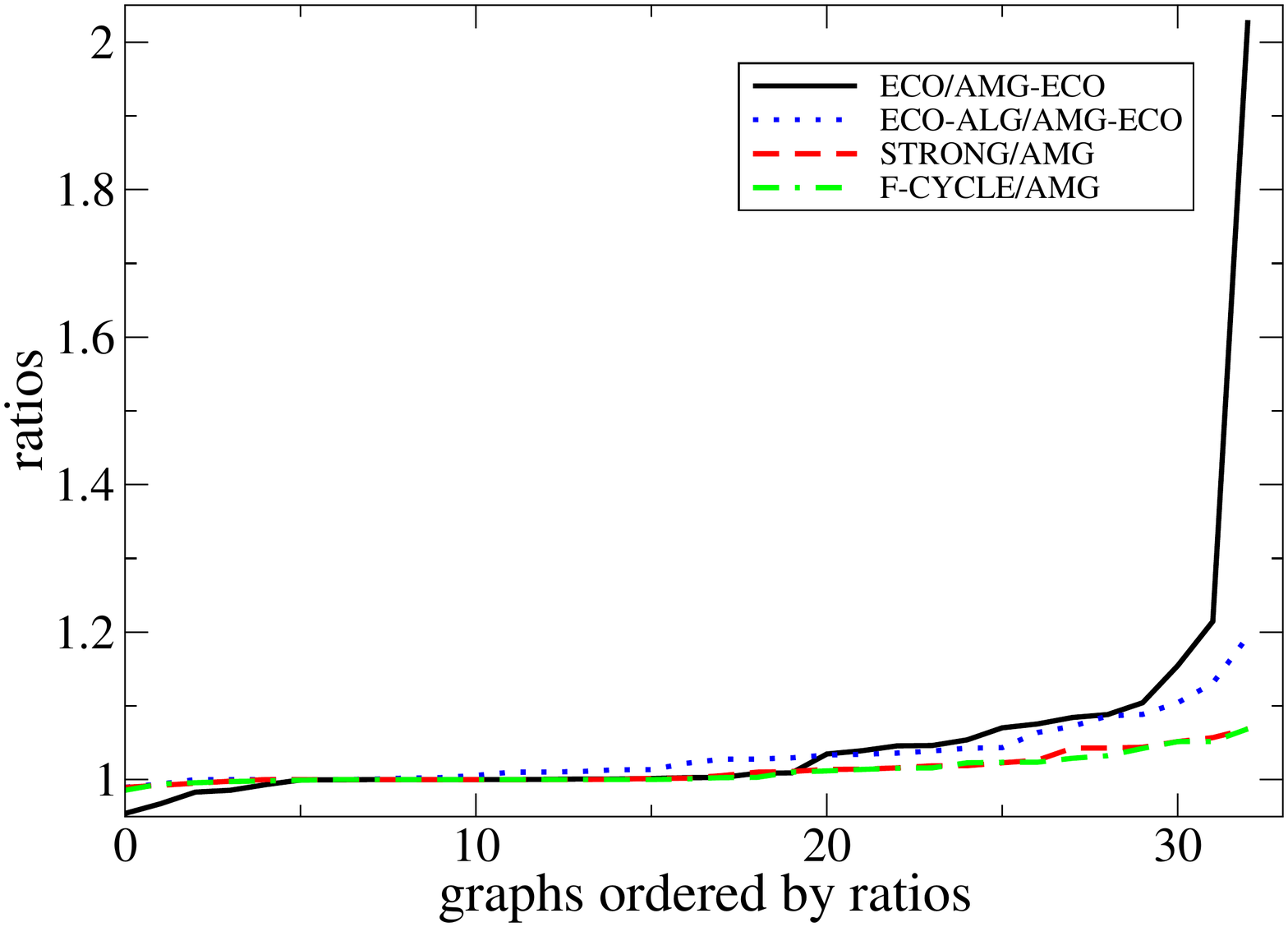}  \hspace{-1cm} 
\includegraphics[width=8cm]{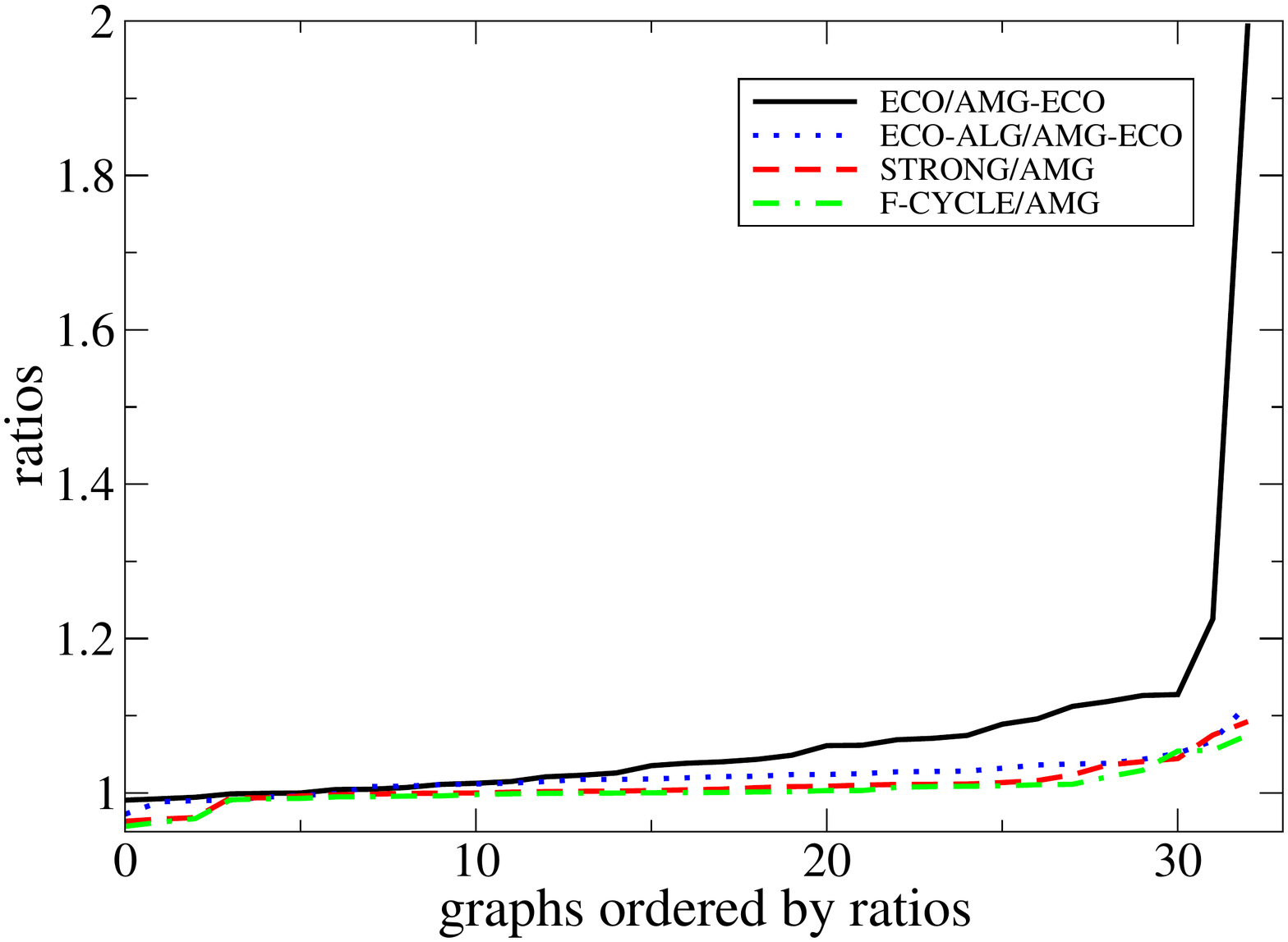}\\
$~$ \hspace{3cm} (a) $k=2$ \hspace{5.5cm} (b) $k=4$ \\
\includegraphics[width=8cm]{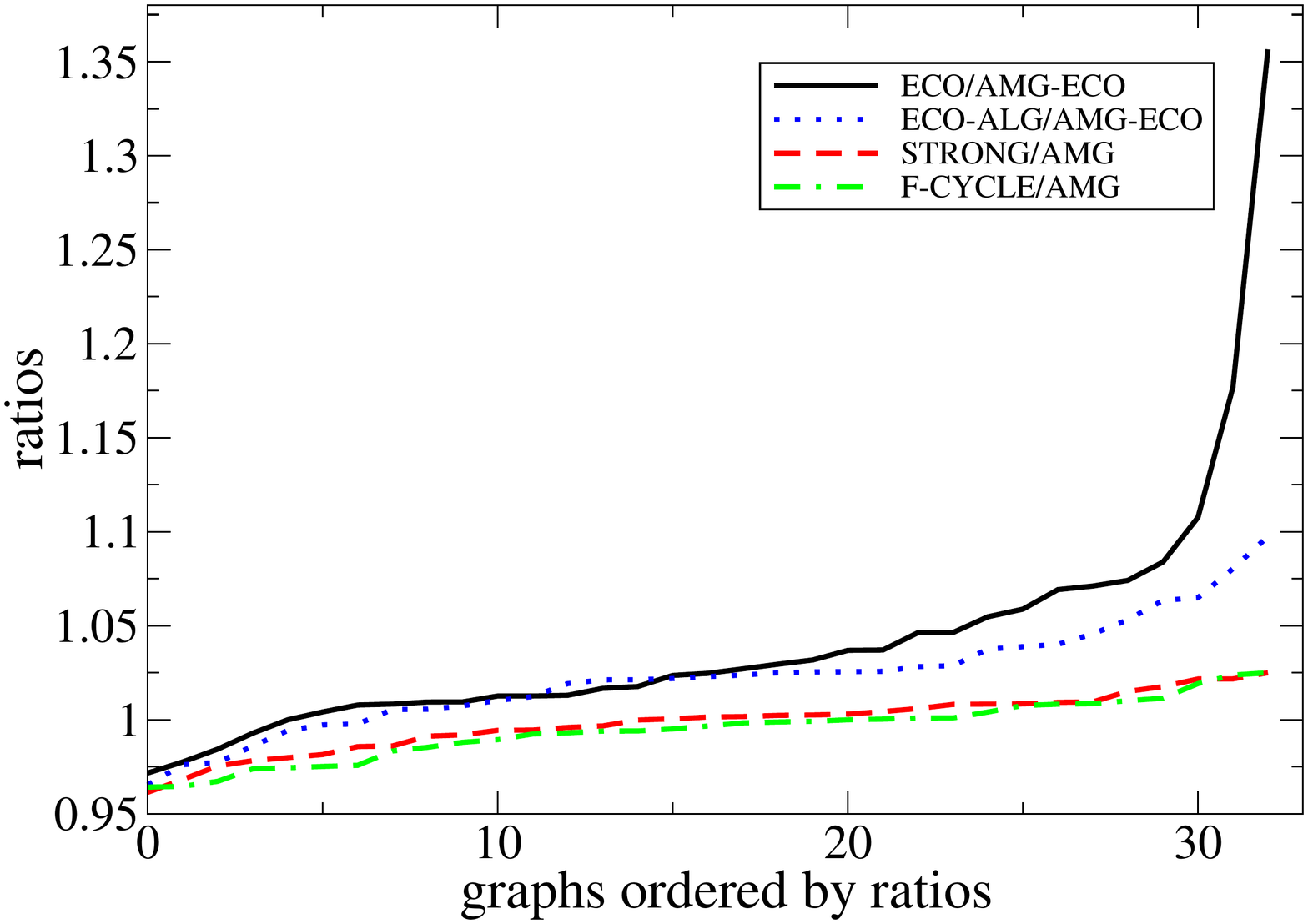}  \hspace{-1cm}
\includegraphics[width=8cm]{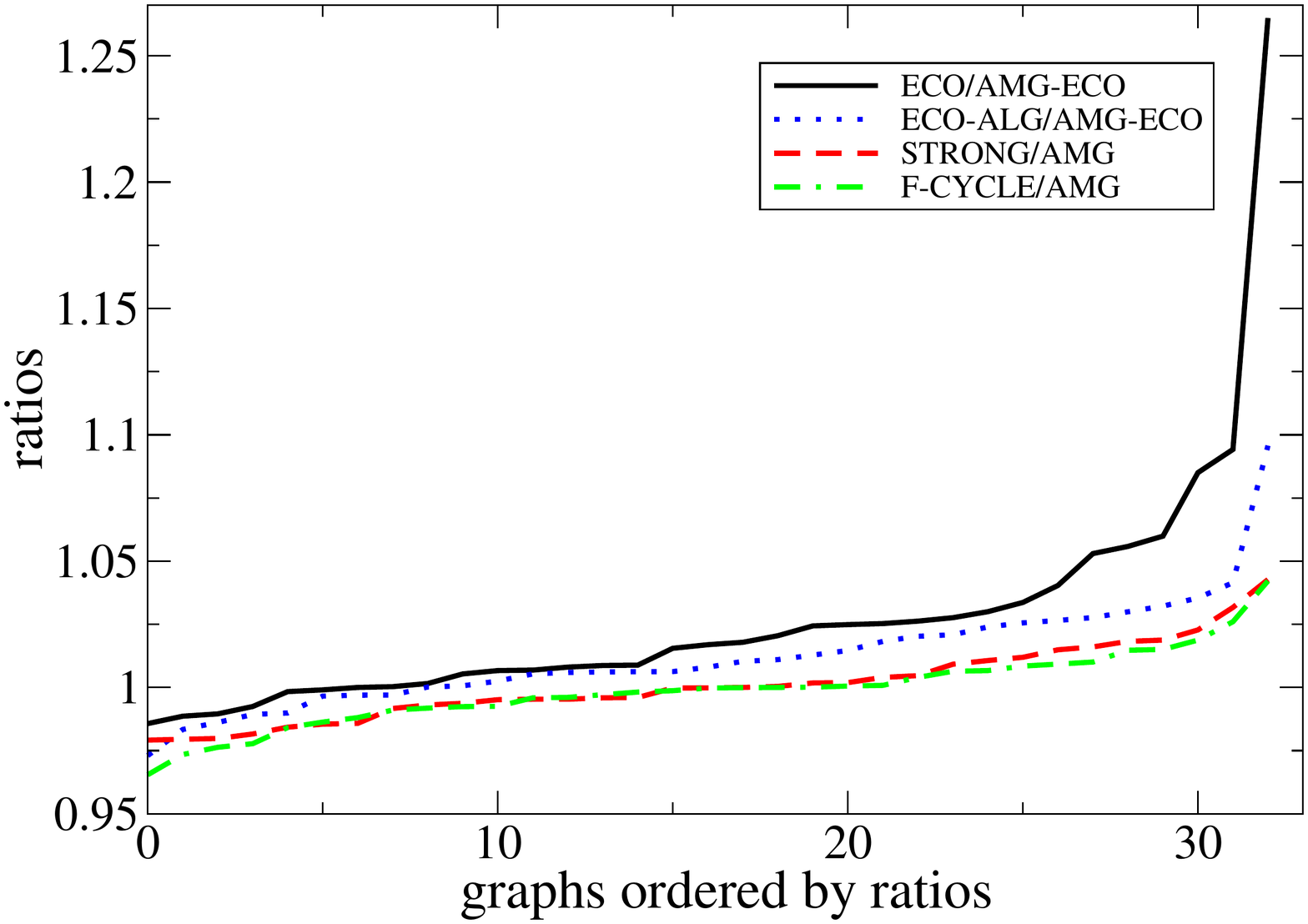}\\
$~$ \hspace{3cm} (c) $k=8$ \hspace{5.5cm} (d) $k=16$ \\
\includegraphics[width=8cm]{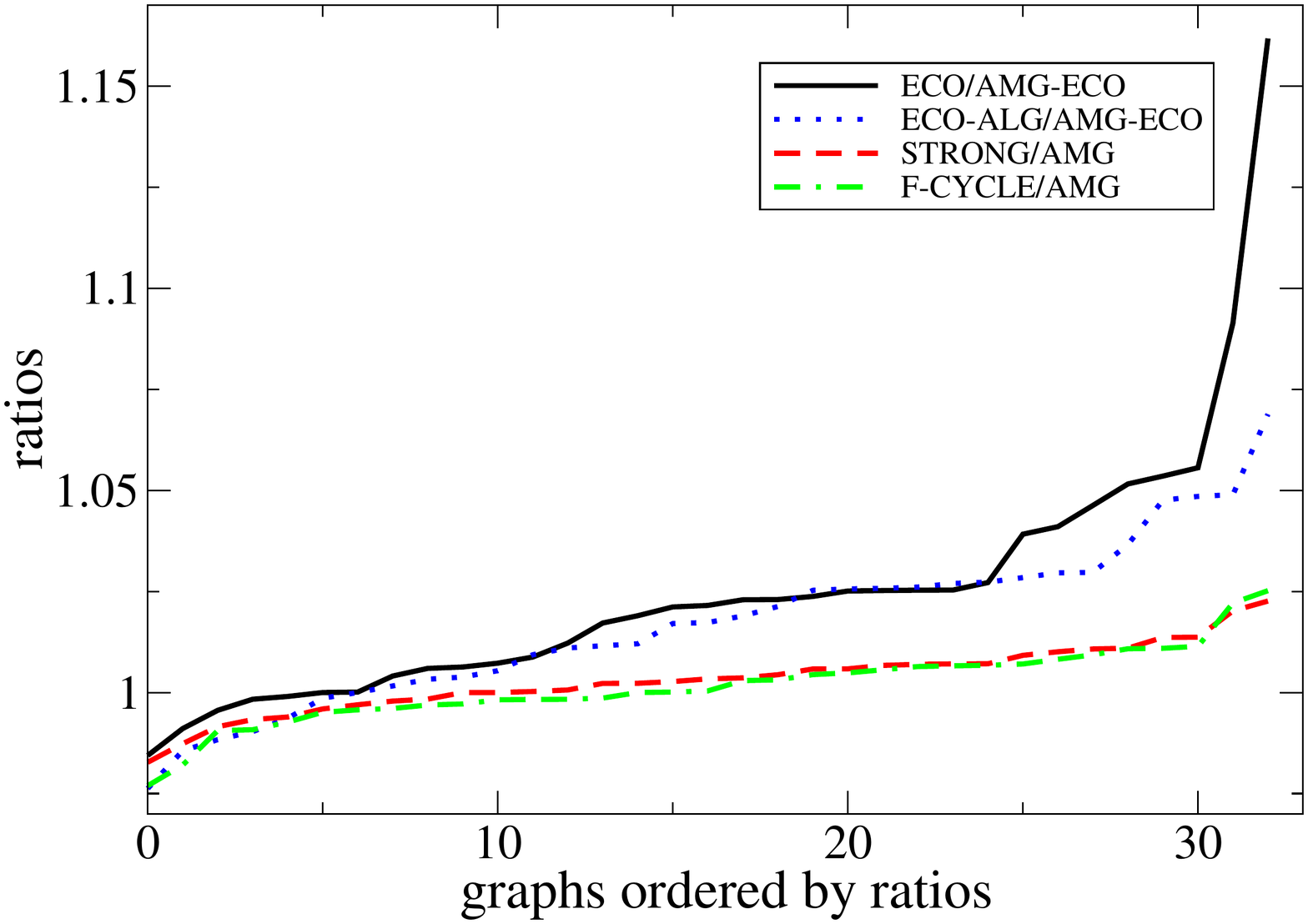}  \hspace{-1cm}
\includegraphics[width=8cm]{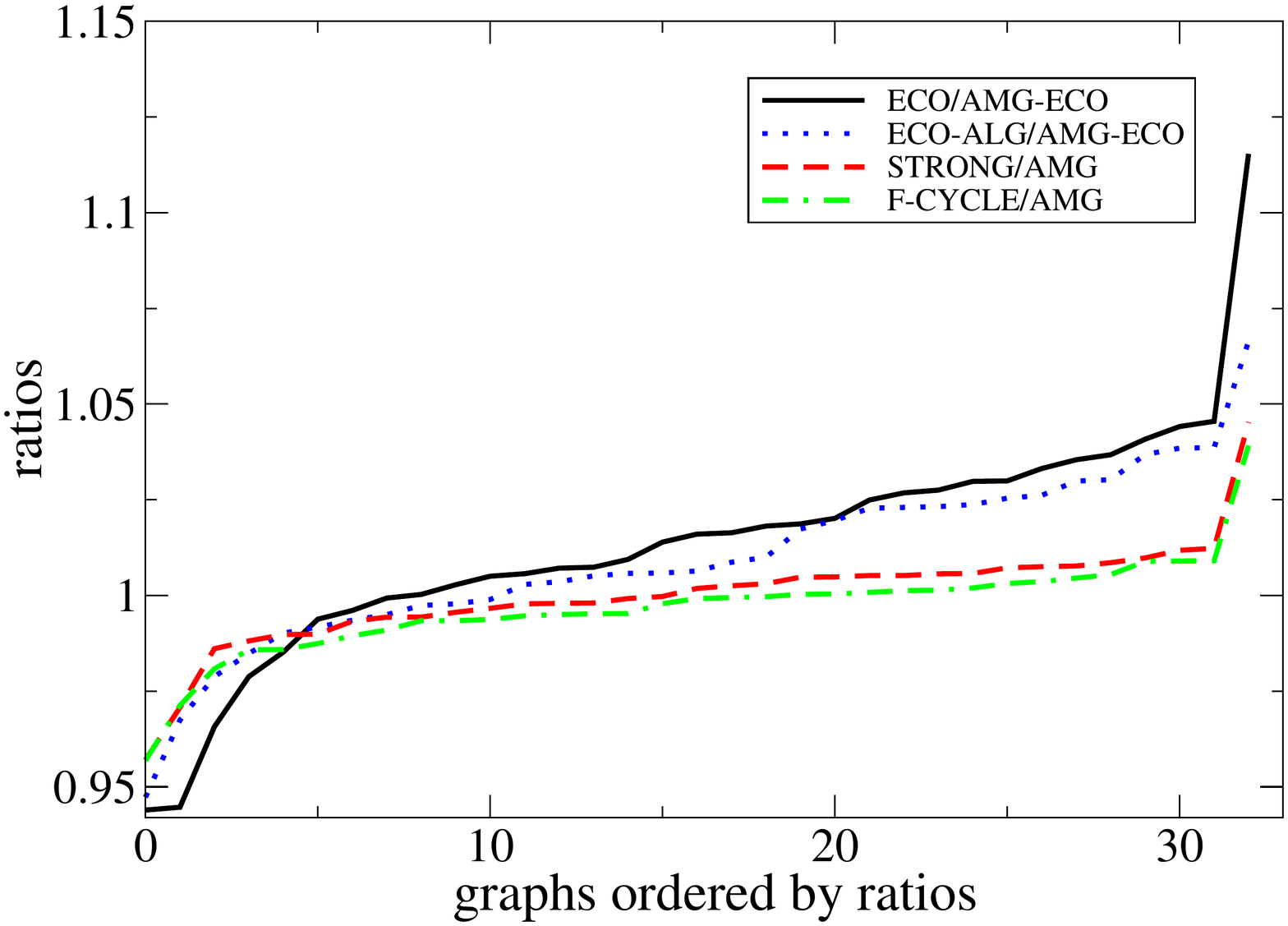}\\
$~$ \hspace{3cm} (e) $k=32$ \hspace{5.5cm} (f) $k=64$ \\
\caption{Comparison of coarsening schemes on Walshaw's benchmark graphs. Figures (a)-(f) contain results of comparison for $k=2$, 4, 8, 16, 32, and 64, respectively. Each figure contains four curves that correspond to ECO/AMG-ECO, ECO-ALG/AMG-ECO, STRONG/AMG, and F-CYCLE/AMG ratios, respectively. Each point on a curve corresponds to the ratio related to one graph.}\label{fig:walshaw} 
\end{figure}

\begin{figure}
\centering
\includegraphics[width=8cm]{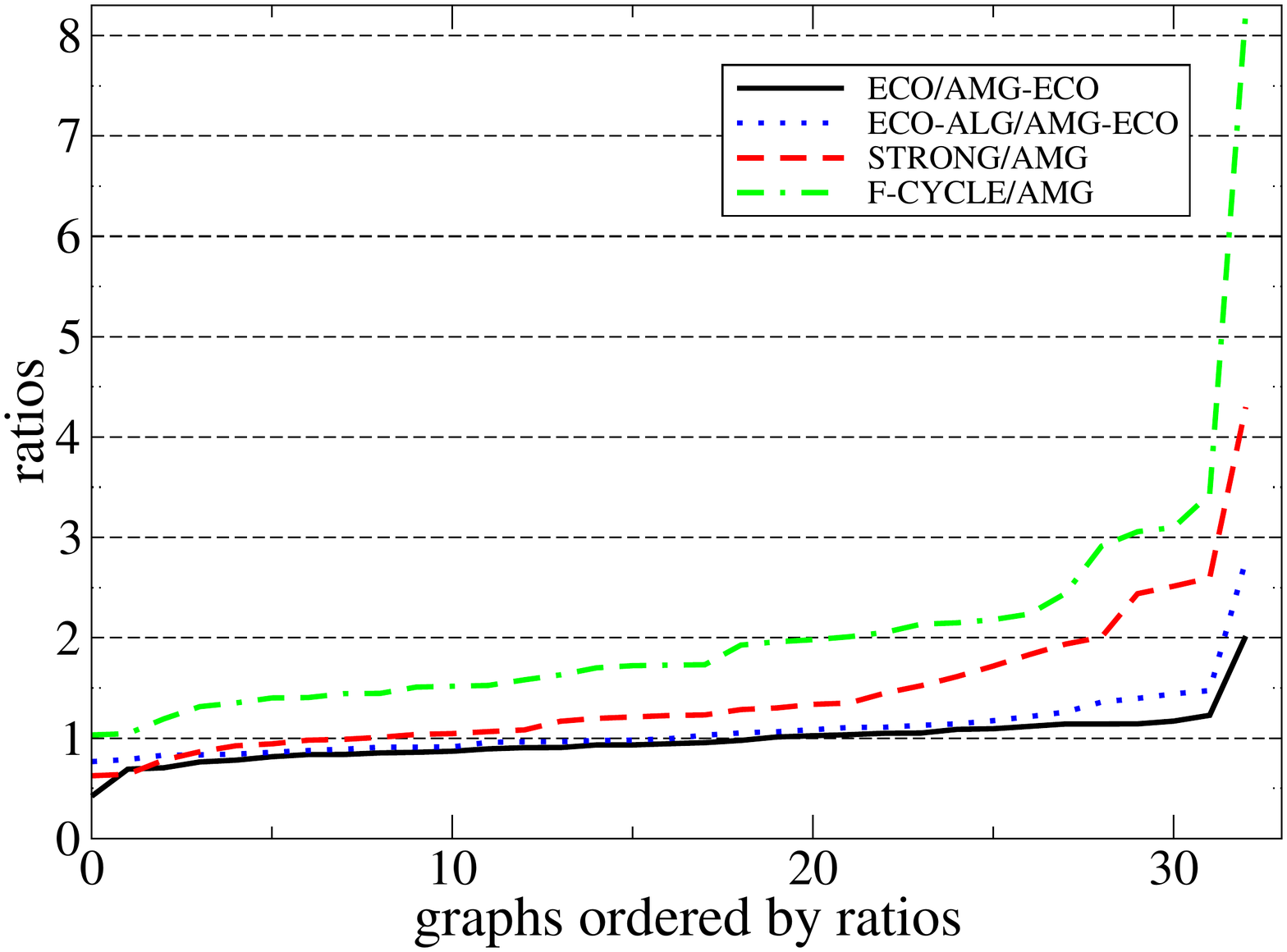}
\caption{Comparison of uncoarsening running time on Walshaw's benchmark graphs for $k=32$. The figure contains four curves that correspond to ECO/AMG-ECO, ECO-ALG/AMG-ECO, STRONG/AMG, and F-CYCLE/AMG ratios, respectively. Each point on curves correspond to the ratio related to one graph.}\label{fig:runtimewalshaw}. 
\vspace*{-.5cm}
\end{figure}

 In Figures \ref{fig:walshaw} we present the results of the comparison of the algorithms on these graphs for different numbers of blocks $k$. 
The horizontal axes represent ordered graphs from the benchmark (however, the ordering itself will be different for each curve). 
The vertical axes are for ratios that represent the comparison of averages of final results for a pair of methods. Each figure contains four curves.
Each curve correspond to a comparison of the following pairs of methods: 
 ECO vs. AMG-ECO, ECO-ALG vs. AMG-ECO, STRONG vs. AMG, and F-CYCLE vs. AMG. Each point on the curves corresponds to the ratio between the average over 10 runs of one method and the average over 10 runs of another method. Each run depends on different random seeds and, thus, can produce different results. For example, the last point at the black solid curve in Figure \ref{fig:walshaw}a has value 2.03, which means that
\[
\frac{\textsf{average}(\textsf{ECO final cut given seed } s_1, \cdots,  \textsf{ ECO final cut given seed } s_{10})}{\textsf{average}(\textsf{AMG-ECO final cut given seed } s_1, \cdots,  \textsf{ AMG-ECO final cut given seed } s_{10})} = 2.03
\]
in experimental series for $k=2$.
A comparison of the running time for uncoarsening phases is presented in Figure \ref{fig:runtimewalshaw}. Each point on the curves in Figure \ref{fig:runtimewalshaw} corresponds to a ratio of uncoarsening running times of two methods. We observed that uncoarsening performance of fast versions (ECO, ECO-ALG, AMG-ECO) are more or less similar to each other. The uncoarsening of a STRONG V-cycle is somewhat slower than AMG because of the density of coarse levels. 
The averages are summarized in Table \ref{tab:walshaw}. Full results are summarized in \cite{hardpart-site}.

\begin{table}[htbp]
\begin{center}
\vspace*{-.25cm}
\begin{tabular}{|l|c|c|c|c|}
\hline
$k$ & ECO/ECO-ALG & ECO-ALG/ECO-AMG & STRONG/AMG & F-CYCLE/AMG \\  \hline
2 & 1.026 & 1.034 & 1.013 & 1.012 \\ 
4 & 1.053 & 1.021 & 1.009 & 1.004 \\ 
8 & 1.019 & 1.023 & 0.998 & 0.995 \\ 
16 & 1.015 & 1.012 & 1.001 & 0.999 \\
32 & 1.008 & 1.017 & 1.003 & 1.002 \\ 
64 & 1.004 & 1.009 & 1.000 & 0.997 \\ \hline
\end{tabular}
\vspace*{.25cm}
\caption{Computational comparison for Benchmark I. Each number corresponds to the ratio of averages of final cuts for pair of methods in the column title and $k$ given in the row.}\label{tab:walshaw}
\vspace*{-1.5cm}
\end{center}
\end{table}
\paragraph{Benchmark II: Scale-free networks.} In scale-free networks the distribution of vertex degrees asymptotically follows the power-law distribution. Examples of such networks include WWW links, social communities, and biological networks. 
These types of networks often contain irregular parts and long-range links (in contrast to Benchmark I) that can confuse both contraction and AMG coarsening schemes. 
Since Walshaw's benchmark doesn't contain graphs derived from such networks, we evaluate our algorithms on 15 graphs collected from \cite{dimacs10,snap}. Full information about these graphs, along with the computational results, is available at \cite{hardpart-site}.
\par
\begin{figure}
\includegraphics[width=9cm]{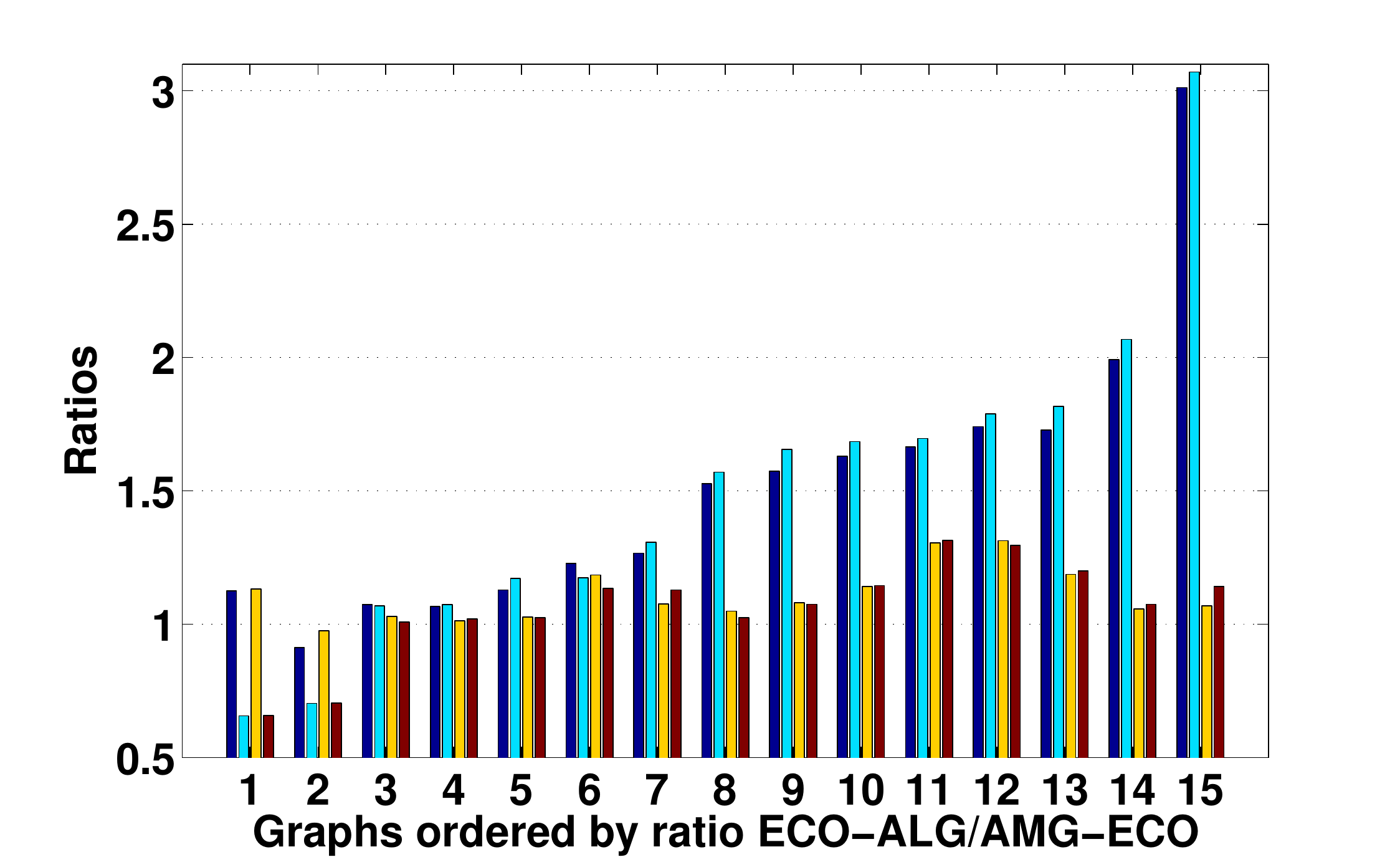}  \hspace{-1cm}
\includegraphics[width=9cm]{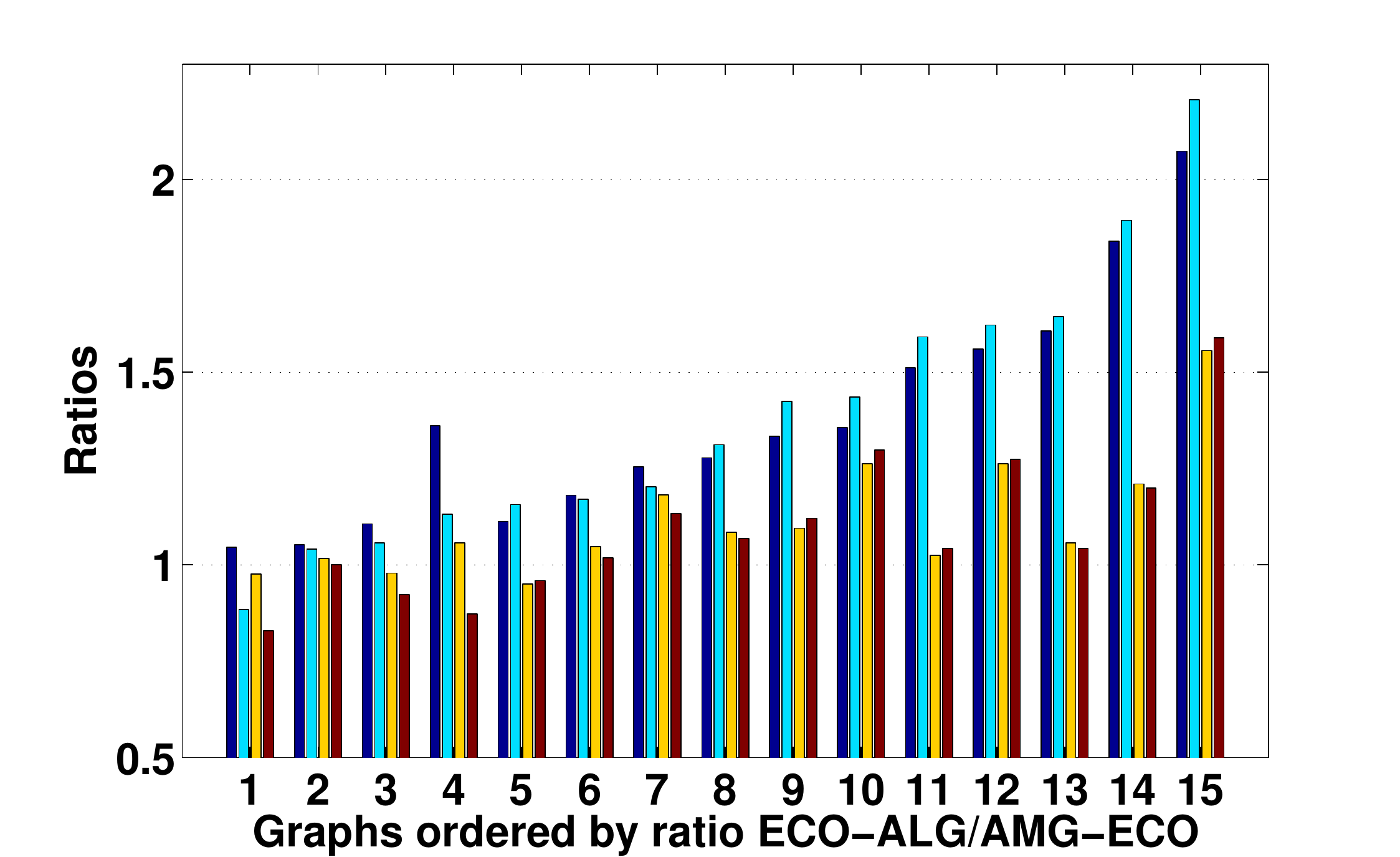}\\
$~$ \hspace{4cm} (a) $k=2$ \hspace{6cm} (b) $k=4$ \\
\includegraphics[width=9cm]{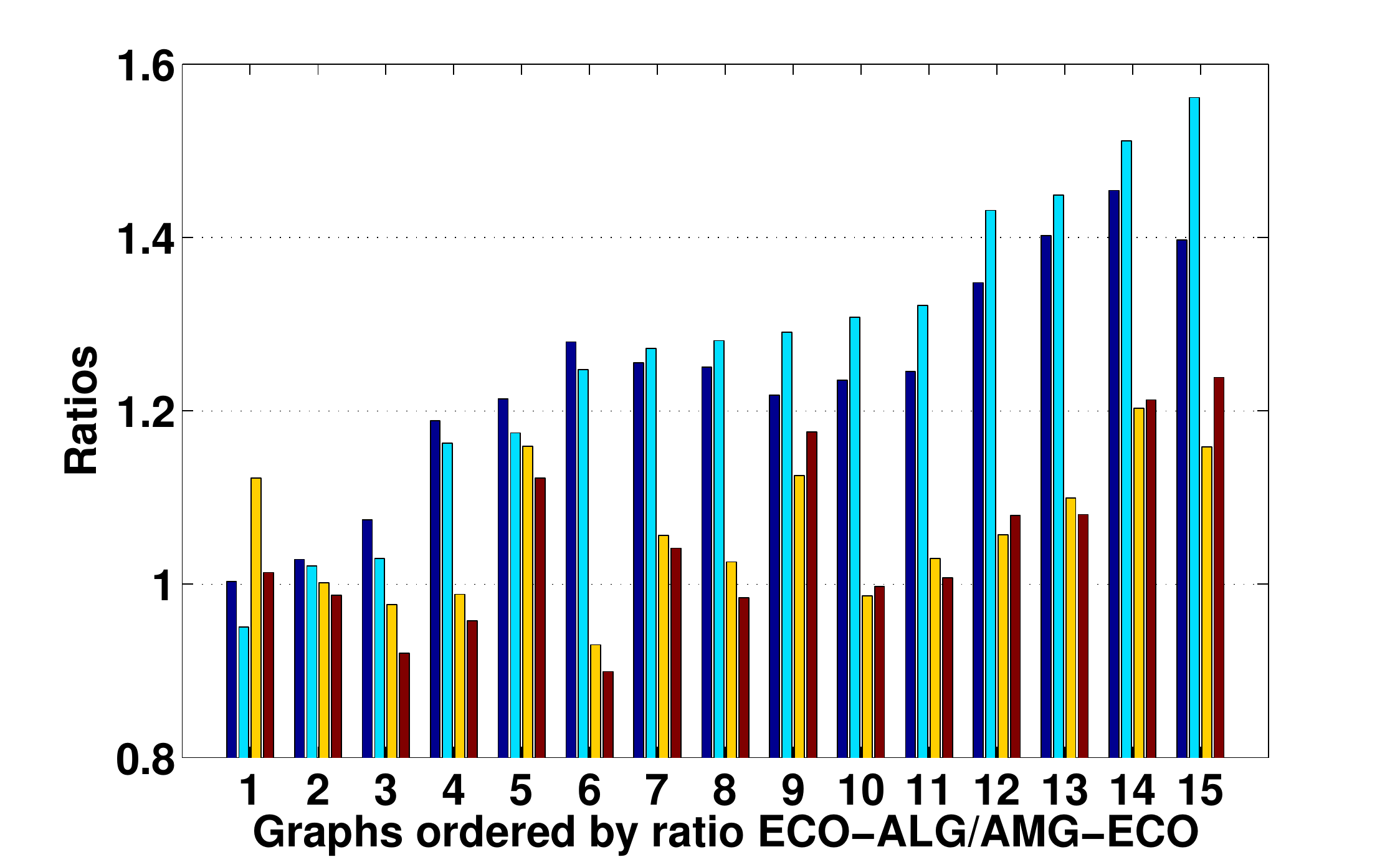}  \hspace{-1cm}
\includegraphics[width=9cm]{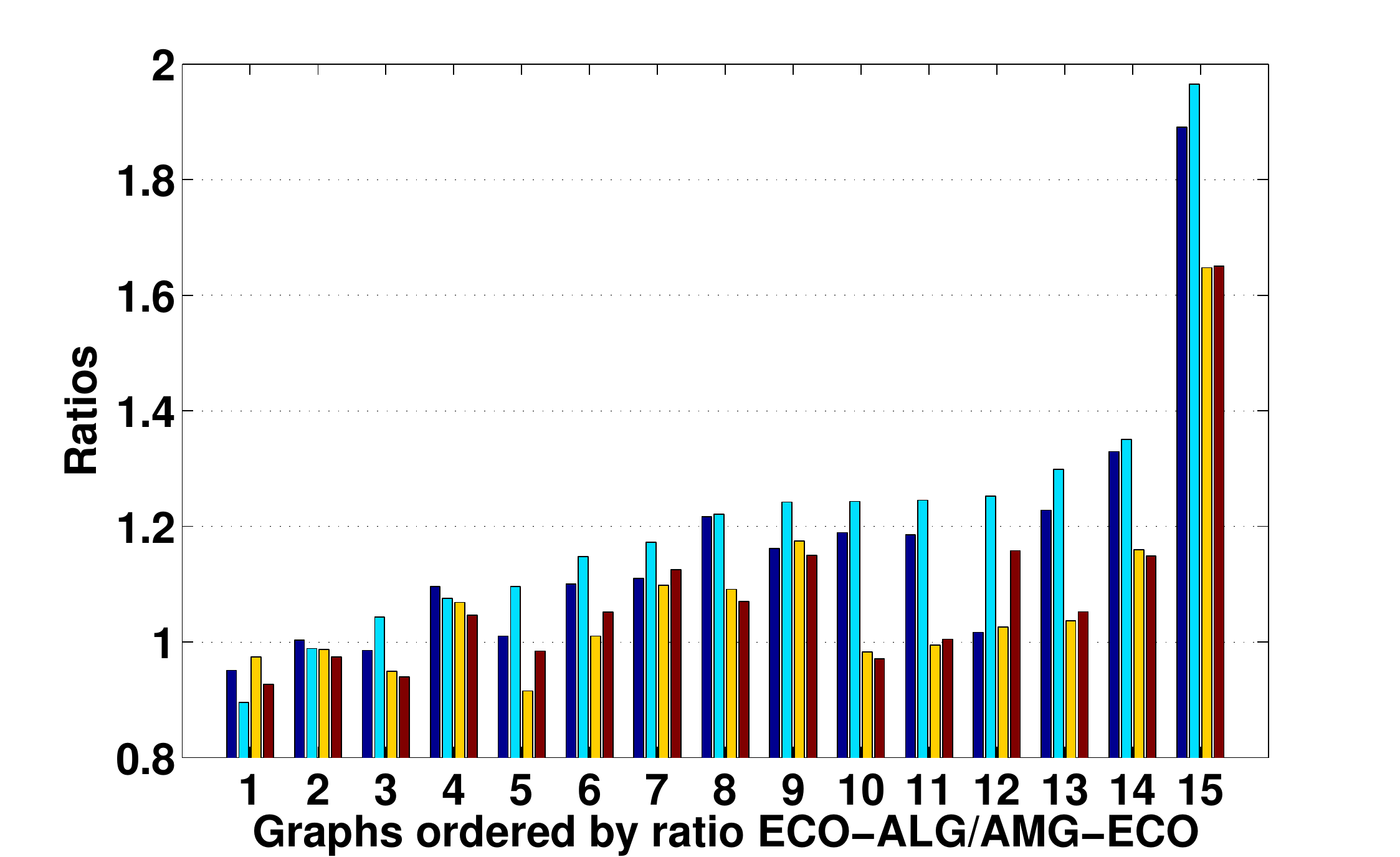}\\
$~$ \hspace{4cm} (c) $k=8$ \hspace{6cm} (d) $k=16$ \\
\includegraphics[width=9cm]{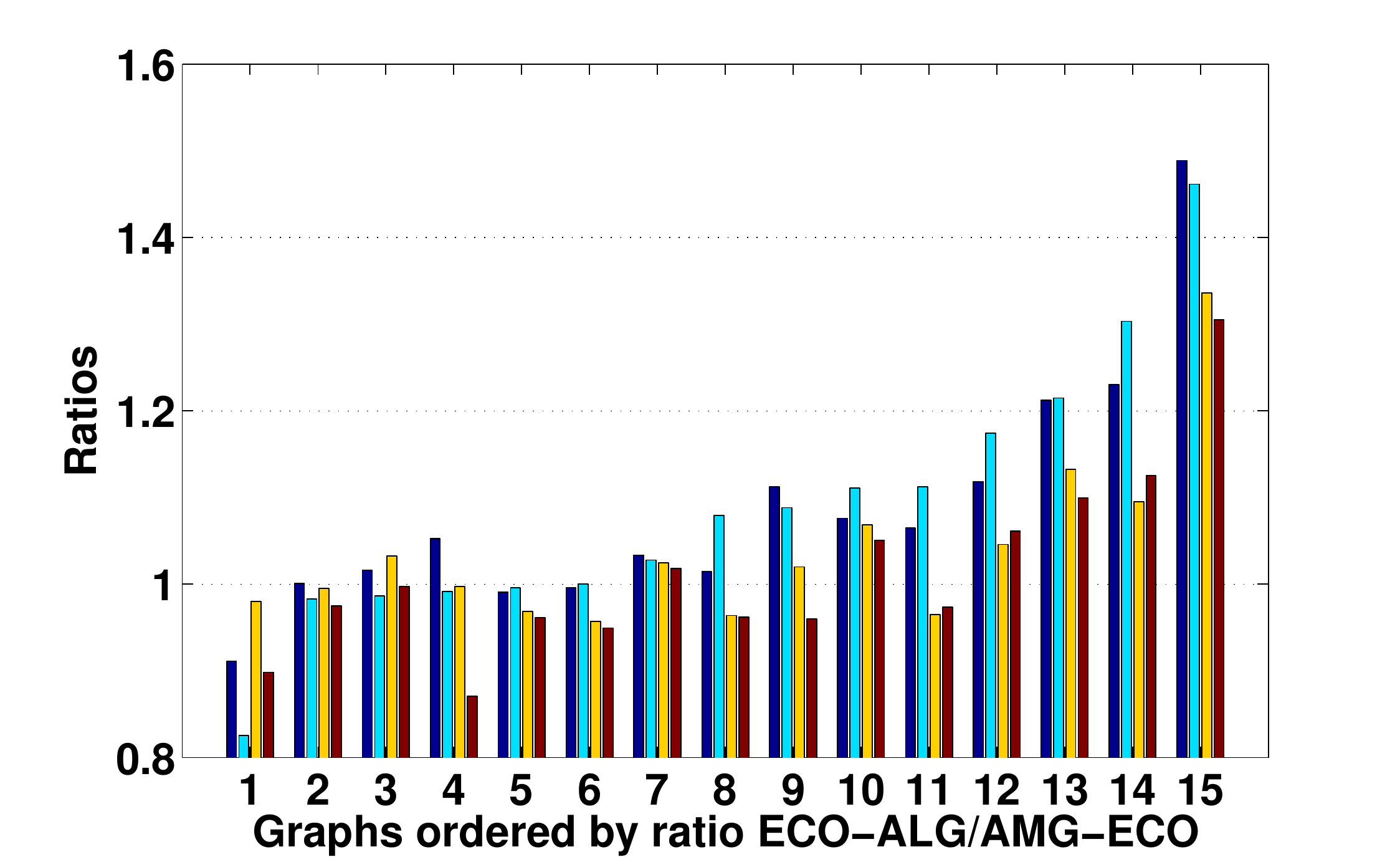}  \hspace{-1cm}
\includegraphics[width=9cm]{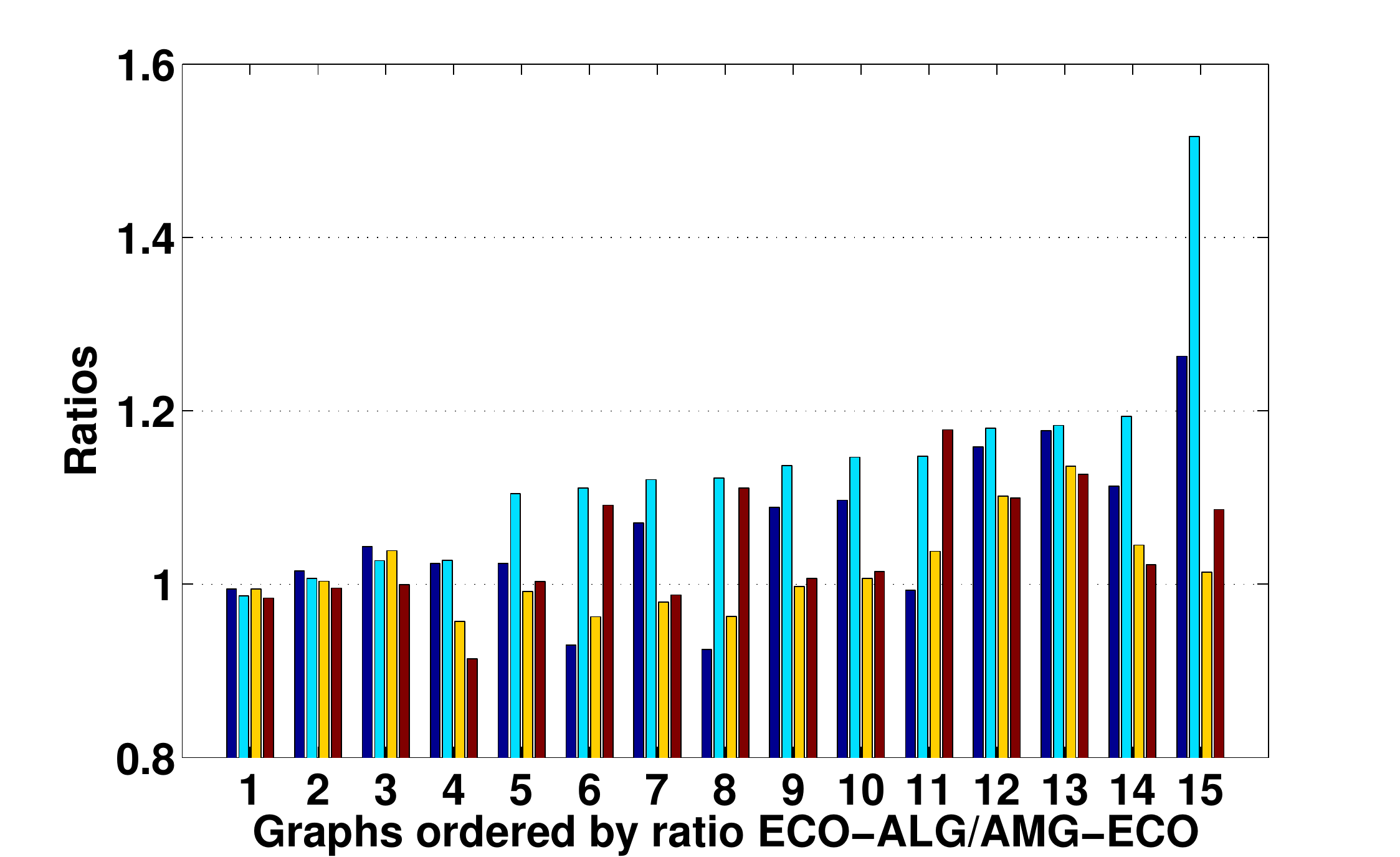}\\
$~$ \hspace{4cm} (e) $k=32$ \hspace{6cm} (f) $k=64$ \\
\caption{Comparison of coarsening schemes on scale-free graphs. Figures (a)-(f) contain results of comparison for $k=2$, 4, 8, 16, 32, and 64, respectively. Each quadruple of bars correspond to one graph. First, second, third and fourth bars represent averages of ratios ECO/AMG-ECO, ECO-ALG/AMG-ECO after refinement, ECO/AMG-ECO, and ECO-ALG/AMG-ECO before refinement, respectively. Three exceptionally high ratios on both Figures are between 2.1 and 3.}\label{fig:scalefree} 
\end{figure}

 The results of the comparison on scale-free graphs are presented in Figure \ref{fig:scalefree}. 
Because of the large running time of the strong configurations on these graphs, we compare only the fast versions of AMG and matching-based coarsening. 
Each figure corresponds to a different number of blocks $k$. The horizontal axes represent graphs from the benchmark. The vertical axes are for ratios that represent comparison of averages of final results for a pair of methods. Each graph corresponds to one quadruple of bars. First, second, third and fourth bars represent averages of ratios ECO/AMG-ECO, ECO-ALG/AMG-ECO after finest refinement, ECO/AMG-ECO, ECO-ALG/AMG-ECO before finest refinement, respectively. As in the previous case the averages are calculated over 10 runs.
\begin{table}[htbp]
\begin{center}
\vspace*{-.25cm}
\begin{tabular}{|l|c|c|c|c|c|}
\hline
 & $\frac{\text{ECO}}{\text{ECO-ALG}}$ & $\frac{\text{ECO}}{\text{ECO-ALG}}$ & $\frac{\text{ECO}}{\text{ECO-ALG}}$ & $\frac{\text{ECO-ALG}}{\text{AMG-ECO}}$ & $\frac{\text{ECO-ALG}}{\text{AMG-ECO}}$ \\ 
 $k$ & quality & full time & uncoarsening time & quality & uncoarsening time \\ \hline
2 & 1.38 & 0.77 & 1.62 & 1.16 & 3.62 \\ 
4 & 1.24 & 1.32 & 1.85 & 1.11 & 2.14 \\ 
8 & 1.15 & 1.29 & 1.45 & 1.07 & 1.94 \\ 
16 & 1.09 & 1.27 & 1.33 & 1.06 & 1.69 \\
32 & 1.06 & 1.18 & 1.23 & 1.00 & 1.60 \\
64 & 1.06 &	1.13 &	1.13 &	1.01 &	2.99 \\ \hline

\end{tabular}
\vspace*{.25cm}
\caption{Computational comparison for scale-free graphs.}\label{tab:social}
\vspace*{-1.5cm}
\end{center}
\end{table}
\paragraph{Benchmark III: Potentially Hard Graphs for Fast $k$-partitioning Algorithms.}  Today multilevel strategies represent one of the most effective and efficient \emph{generic} frameworks for solving the graph partitioning problem on large-scale graphs. 
The reason is obvious: given a successful global optimization technique $X$ for this problem, one can consider applying it locally by introducing a chain of subproblems along with fixed boundary conditions. 
 Given this and if the coarsening preserves the structural properties of the graph well enough 
, the multilevel heuristic can behave better and work faster
than a direct global application of the optimization technique $X$. 
Examples of such combinations include FM/KL, spectral and min-cut/max-flow techniques with multilevel frameworks. When can the multilevel framework produce low quality results?
\par We present a simple strategy for checking the quality of multilevel schemes. 
To construct a potentially hard instance for gradual multilevel projections, we consider a mixture of graphs that are weakly connected with each other. 
These graphs have to possess different structural properties (such as finite-element faces, power-law degree distribution, and density) to ensure nonuniform coarsening and mutual aggregation of well-separated graph regions. Such mixtures of structures may have a twofold effect. First, they can force the algorithm to contract incorrect edges; and second, they can attract a "too strong" refinement to reach a local optimum, which can contradict better local optimums at finer levels. The last situation has been observed in different variations also in multilevel linear ordering algorithms \cite{safro2005}.
In other words, the uneven aggregation with respect to the scales (not to be confused with uneven sizes of clusters) can lead refinement algorithms to wrong local attraction basins. Examples of graphs that contain such mixtures of structures include multi-mode networks \cite{multimode} and logistics multi-stage system networks \cite{stock2006strategic}. In general, such graphs can be difficult not only to the multilevel algorithms.

\par We created a benchmark (available at \cite{dimacs10}) with potentially hard mixtures. Each graph in this benchmark represents a star-like structure of different graphs $S_0,\dots,S_t$. Graphs $S_1,\dots,S_t$ are weakly connected to the center $S_0$ by random edges. Since all the constituent 
graphs are sparse, a faster aggregation of them has been achieved by adding more than one random edge to each boundary node. The total number of edges between each $S_i$ and $S_0$ was less than 3\% out of the total number of edges in $S_i$. We considered the mixtures of the following structures: social networks, finite-element graphs, VLSI chips, peer-to-peer networks, and matrices from optimization solvers.
\par
\begin{figure}
\includegraphics[width=9cm]{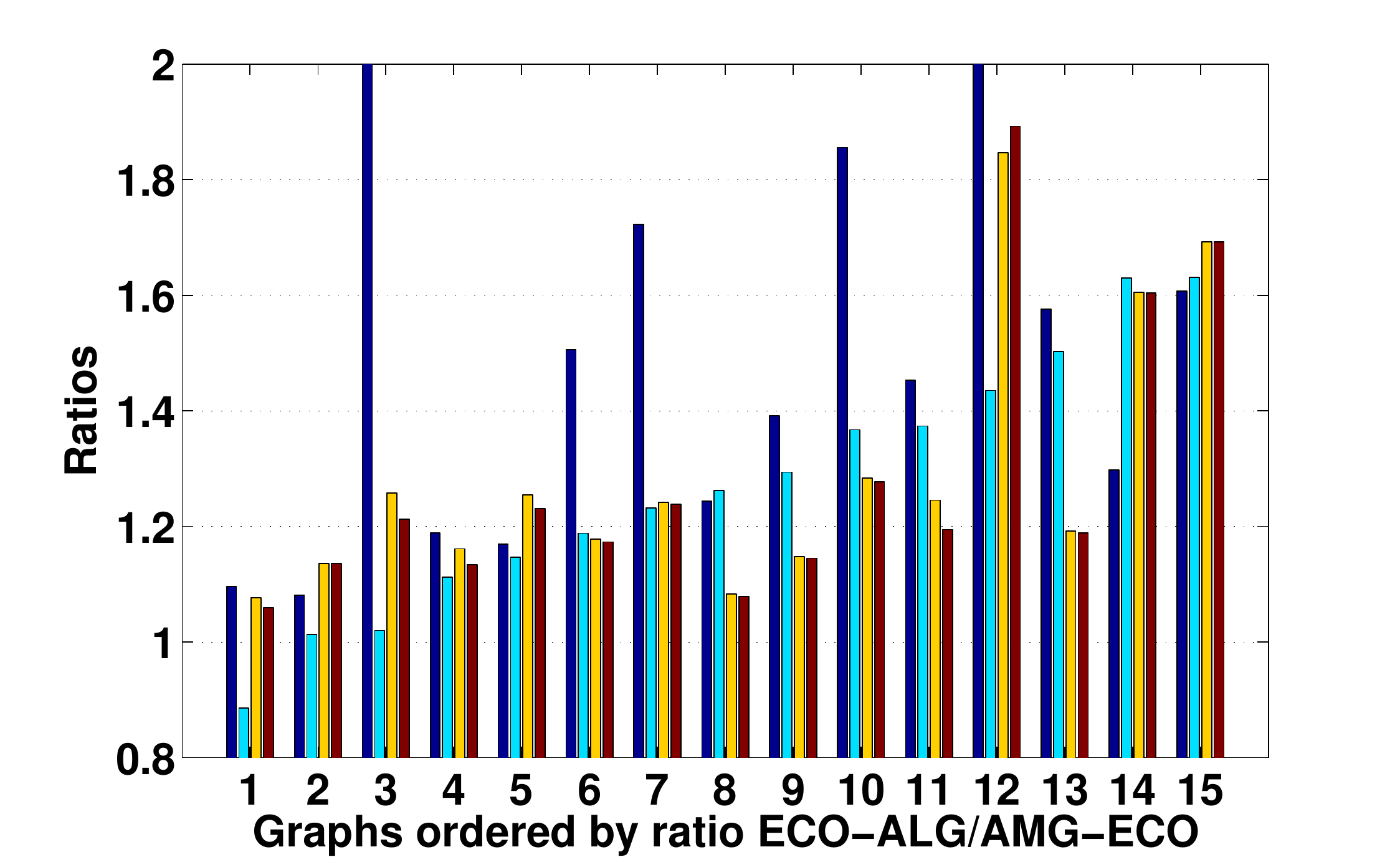}  \hspace{-1cm}
\includegraphics[width=9cm]{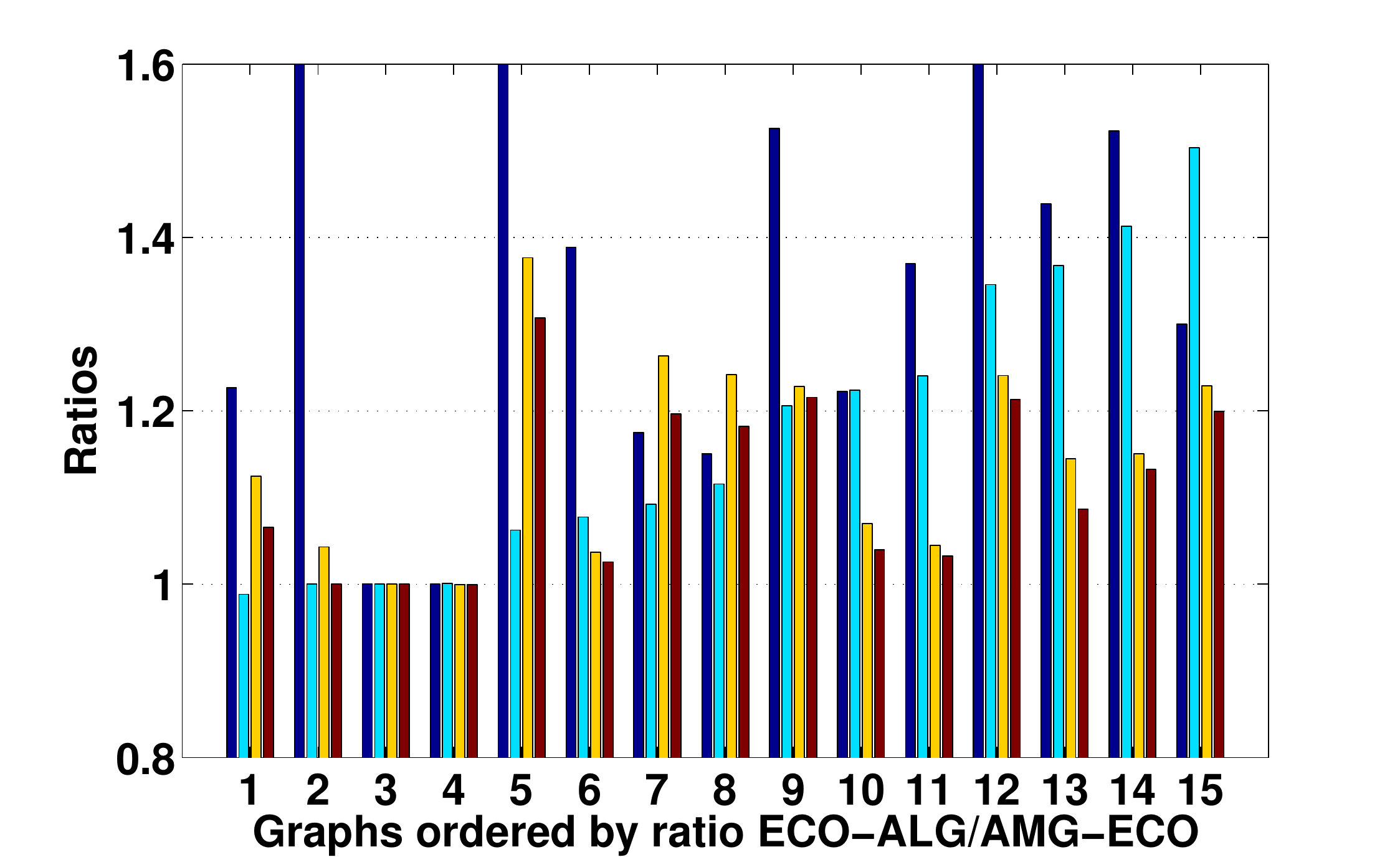}\\
$~$ \hspace{4cm} (a) $k=2$ \hspace{4cm} (b) $k=2$, before last refinement \\
\includegraphics[width=9cm]{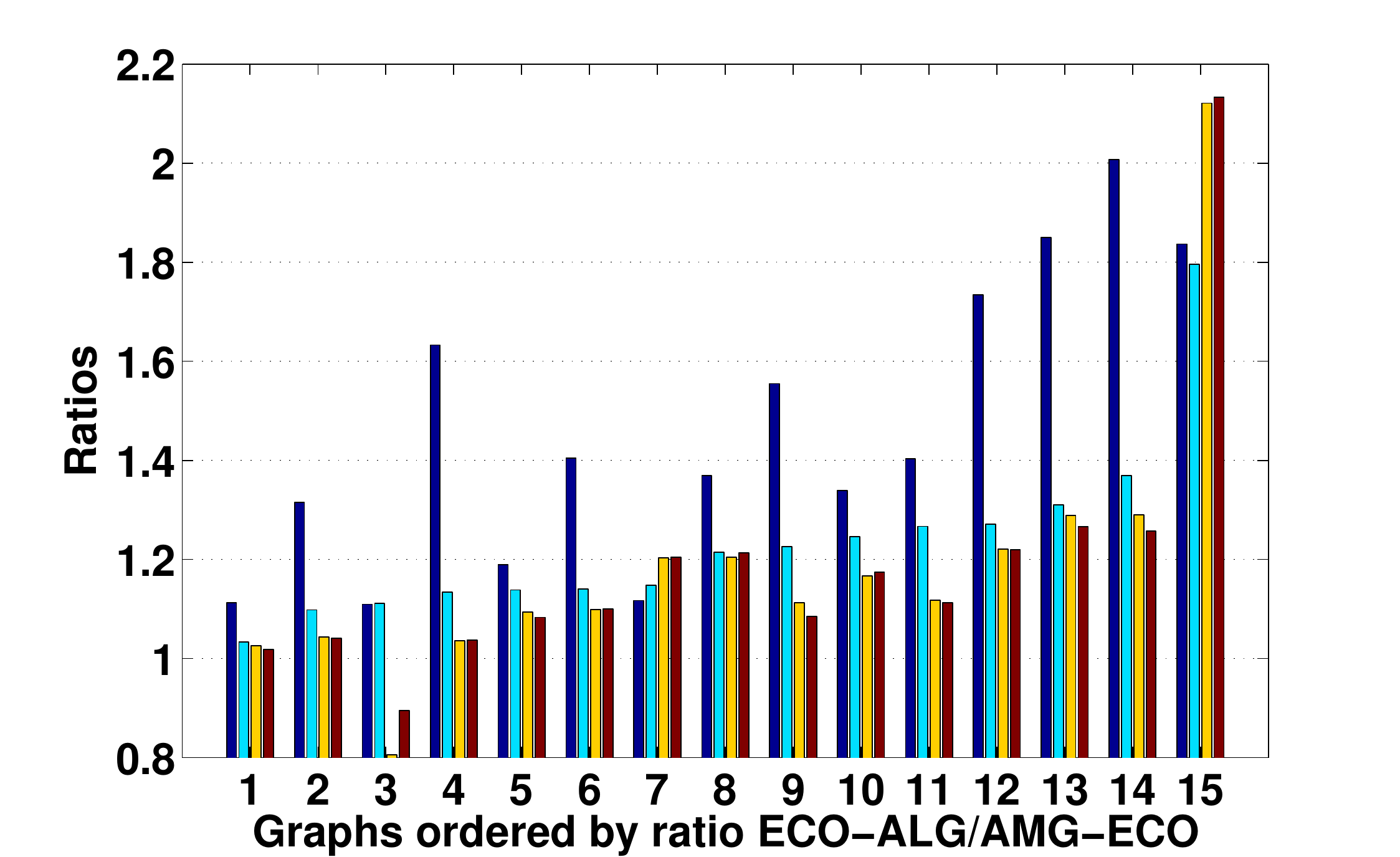}  \hspace{-1cm}
\includegraphics[width=9cm]{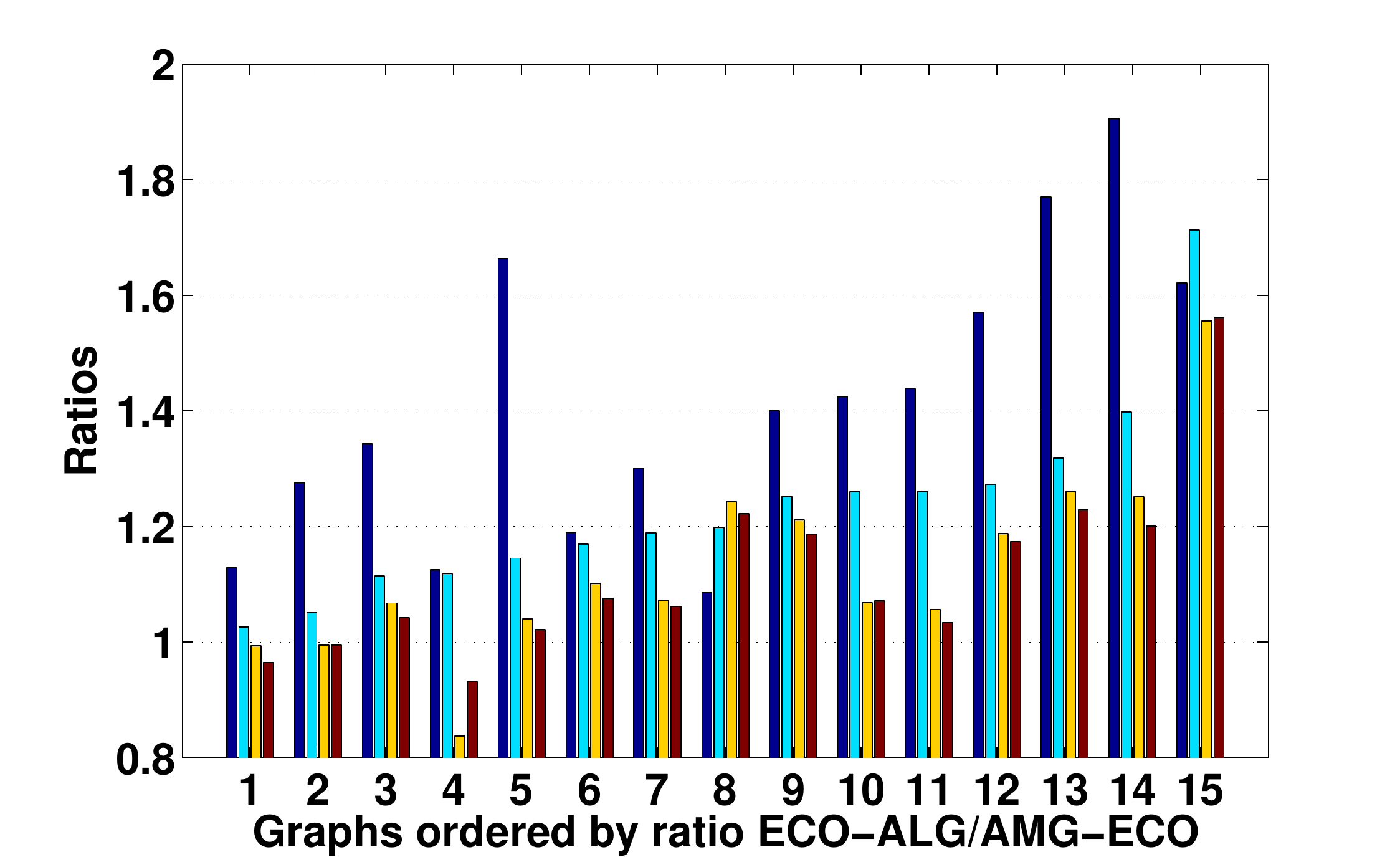}\\
$~$ \hspace{4cm} (c) $k=4$ \hspace{4cm} (d) $k=4$, before last refinement\\
\includegraphics[width=9cm]{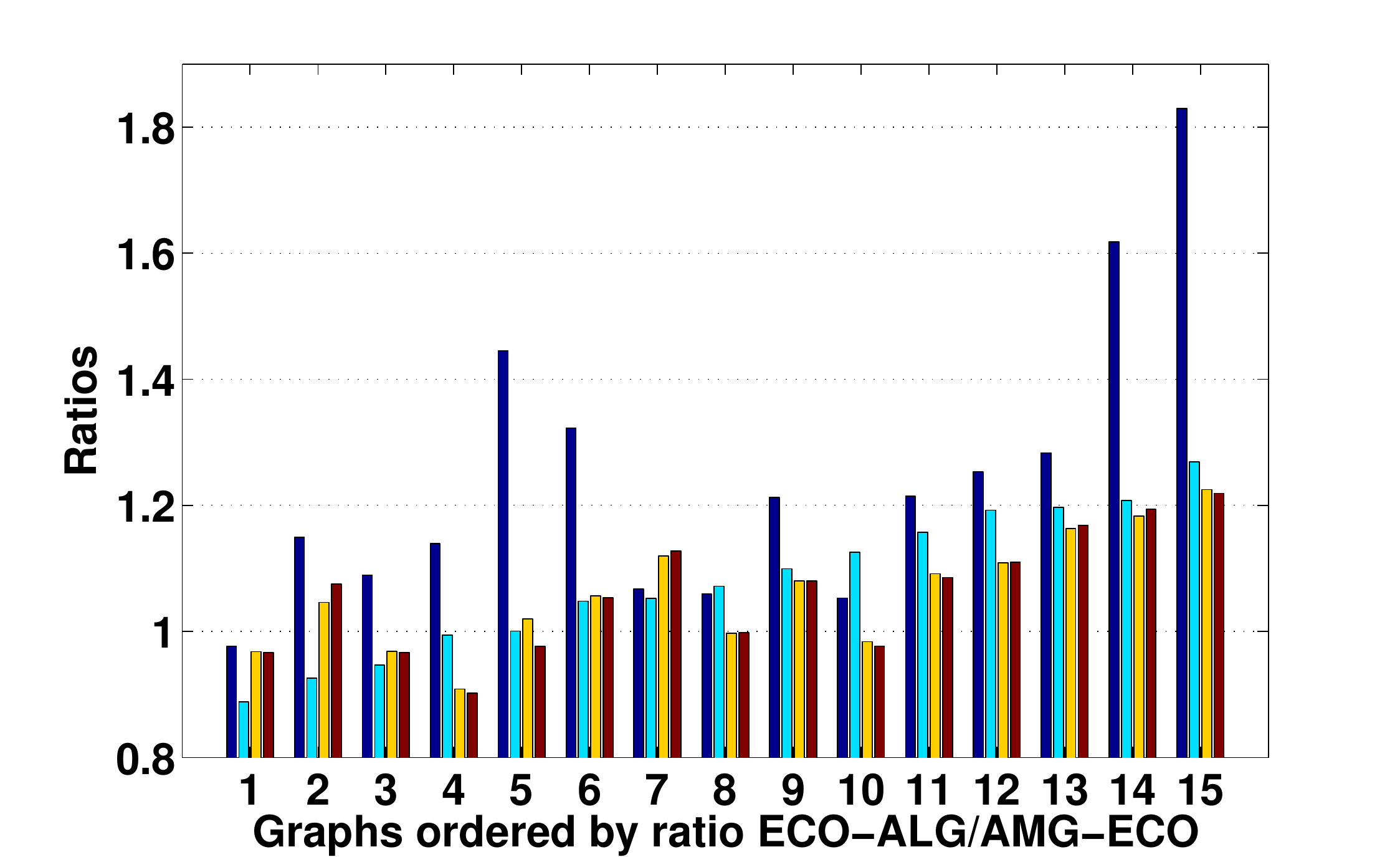}  \hspace{-1cm}
\includegraphics[width=9cm]{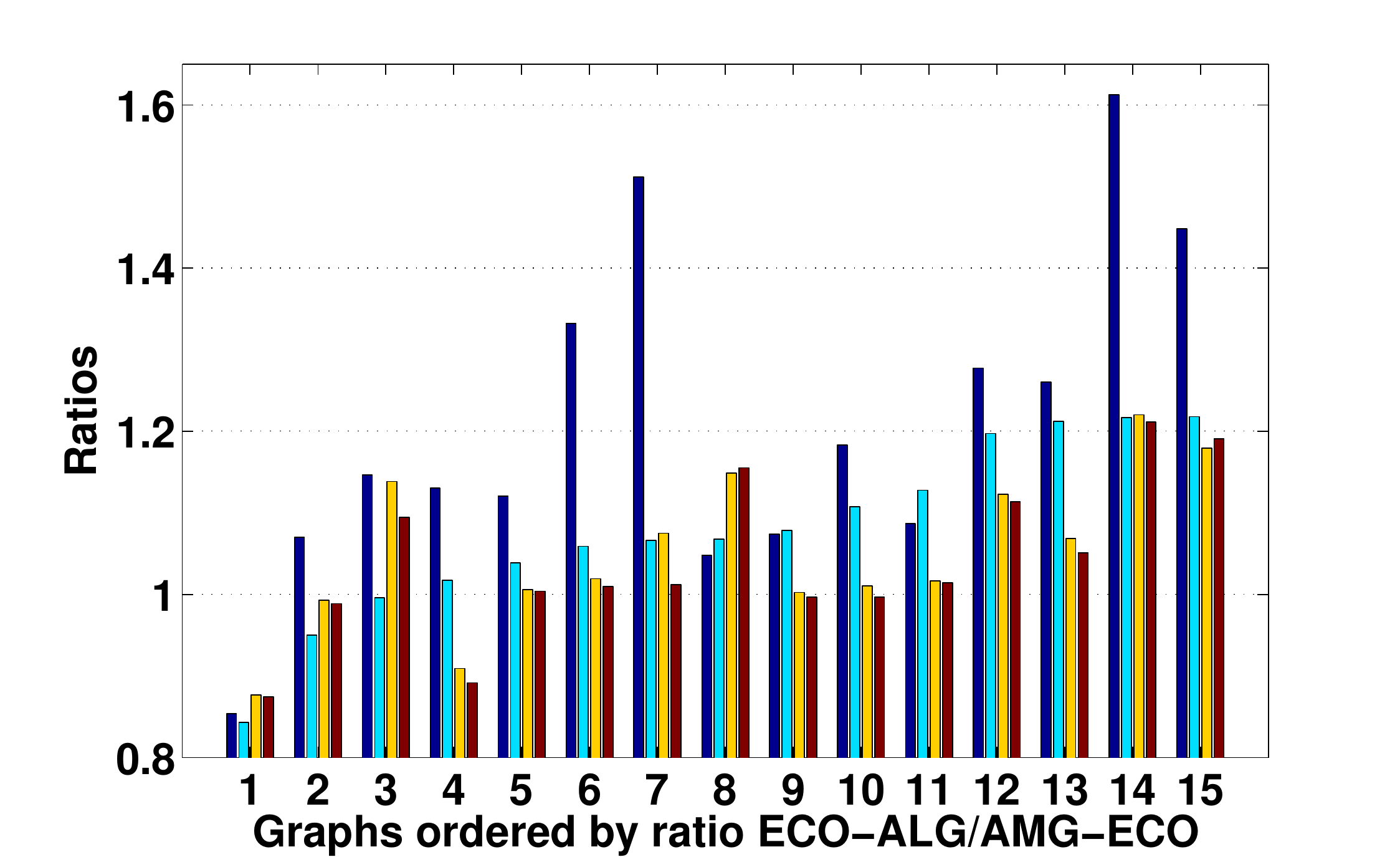}\\
$~$ \hspace{4cm} (e) $k=8$ \hspace{4cm} (f) $k=8$, before last refinement\\
\caption{Comparison of coarsening schemes on hard examples. Figures (a,c,e) contain results of comparison before applying finest level refinement. Figure (b,d,f) contain results of comparison of final results. Each quadruple of bars correspond to one graph. First, second, third and fourth bars represent averages of ratios ECO/AMG-ECO, ECO-ALG/AMG-ECO, STRONG/AMG, and F-cycle/AMG, respectively. Four exceptionally high ratios on both Figures are between 3.5 and 5.7.}\label{fig:hardgraphs}
\end{figure}

 The comparison on this benchmark is demonstrated in Figure \ref{fig:hardgraphs}. Each graph corresponds to one quadruple of bars. The first, second, third and the fourth bar represent averages over 10 ratios of ECO/AMG-ECO, ECO-ALG/AMG-ECO, STRONG/AMG, and F-cycle/AMG, respectively. In almost all experiments we observed that introduction of algebraic distance as a measure of connectivity plays a crucial role in both fast versions AMG-ECO and ECO-ALG since it
helps to separate the subgraphs and postpone their aggregation into one mixture. We also observe that both fast and slow AMG coarsenings almost always lead to better results. Note that in contrast to Benchmarks I and  II, the uncoarsening of ECO-ALG is significantly faster than that of AMG-ECO.

\begin{figure}[t]
\vspace*{-.5cm}
\centering
\includegraphics[width=8cm]{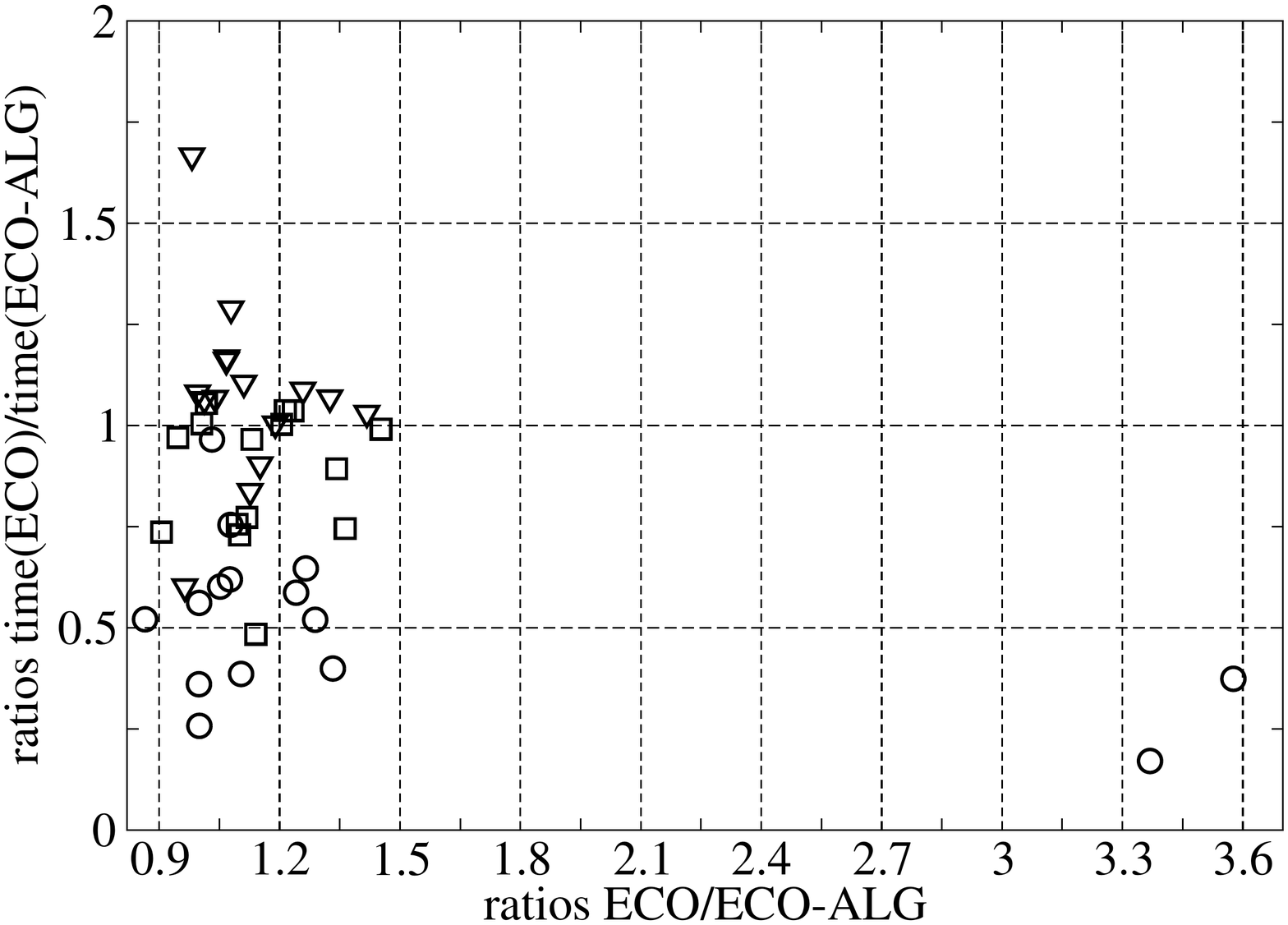} \hspace{-.5cm}
 \includegraphics[width=8cm]{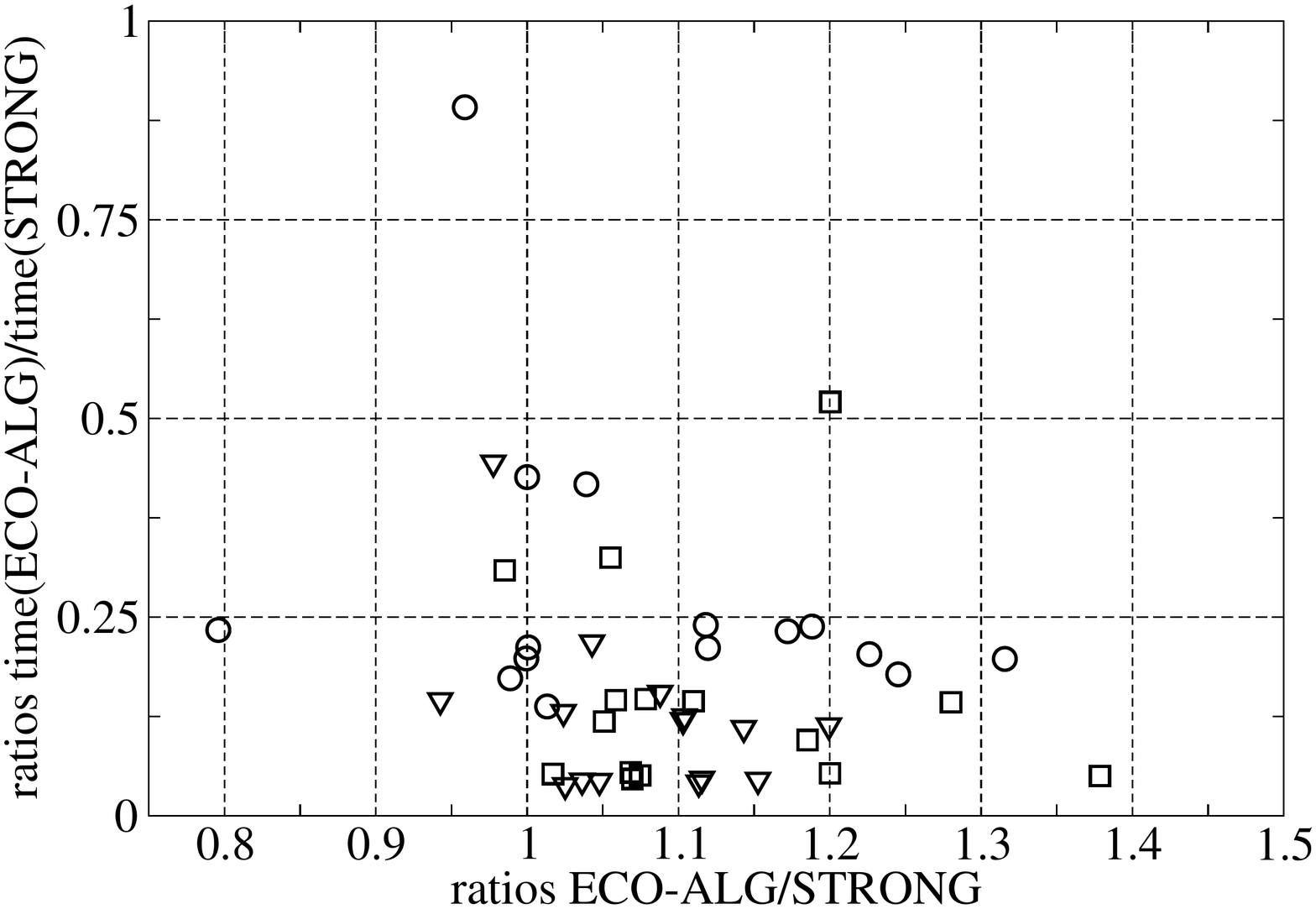}
                        \vspace*{-.5cm}
\caption{Benchmark III: Trade-off between changes in the objectives (horizontal axis) and those in the running times (vertical axis) on Benchmark III. Data points for $k=2$, 4, and 8 are represented by circles, squares, and triangles, respectively. Average ratios are calculated each over 10 runs similarly to previous figures. The left and right figures describe the comparison for ECO vs. ECO-ALG and ECO-ALG vs. STRONG configurations, respectively.}\label{fig:hard-tradeoff}
                        \vspace*{-.5cm}
\end{figure}

\begin{table}[t]
        
\begin{center}
\vspace*{-.25cm}
\begin{tabular}{|l|c|c||c|c||c|c||c|}
\hline
 & $\frac{\text{ECO}}{\text{ECO-ALG}}$ & $\frac{\text{ECO}}{\text{ECO-ALG}}$ & $\frac{\text{ECO-ALG}}{\text{AMG-ECO}}$ & $\frac{\text{ECO-ALG}}{\text{AMG-ECO}}$ & $\frac{\text{STRONG}}{\text{AMG}}$ & $\frac{\text{STRONG}}{\text{AMG}}$ & $\frac{\text{F-CYCLE}}{\text{AMG}}$ \\ 
 & quality & full & quality & uncoarsening & quality & uncoarsening  & quality \\ 
$k$ &         &   time   &         & time         &         & time & \\ \hline
2 & 1.42 & 0.51 & 1.18 & 0.55 & 1.15 & 2.11 & 1.11 \\ 
4 & 1.15 & 0.88 & 1.23 & 0.64 & 1.13 & 1.69 & 1.12 \\ 
8 & 1.12 & 1.08 & 1.08 & 0.98 & 1.05 & 1.37 & 1.04 \\ 
\hline
\end{tabular}
\vspace*{.25cm}
\caption{Computational comparison for potentially hard graphs. 
}\label{tab:hard}
\vspace*{-1.25cm}

\end{center}
\end{table}
\paragraph{Role of the algebraic distance.} In this work the importance of the algebraic distance as a measure of connectivity strength for graph partitioning algorithms has been justified in almost all experimental settings. In particular, the most significant gap was observed between ECO and ECO-ALG (see all benchmarks), versions which confirms preliminary experiments in \cite{chen-safro-algdist-full}, where the algebraic distance has been used at the finest level only. The price for improvement in the quality is the additional running time for Jacobi over-relaxation, which can be implemented by using the most suitable (parallel) matrix-vector multiplication method. However, in cases of  strong configurations and/or large irregular instances, the difference in the running time becomes less influential as it is not comparable to the amount of work in the refinement phase. 
For example, for the largest graph in Benchmark I (auto, $|V|=448695$, $|E|=3314611$) the ECO coarsening is approximately 10 times faster than that in the ECO-ALG; but for both configurations when $k=64$, it takes less than $3\%$ of the total time.
Note that for irregular instances from Benchmark II, already starting $k=4$ the total running time for ECO becomes bigger than in ECO-ALG (see Table \ref{tab:social}).
More examples of trade-off between changes in the objectives and those in the running times on Benchmark III are presented in Figure \ref{fig:hard-tradeoff}.
\paragraph{Does AMG coarsening help?} The positive answer to this question is given mostly by Benchmarks II and III, which contain relatively complex instances (Tables \ref{tab:social} and \ref{tab:hard}). 
On Benchmark III we have demonstrated that the AMG configuration is superior to F-CYCLE, which runs significantly longer. 
This result is in contrast to Benchmark I, in which we did not observe any particular class of graphs that corresponded to stable significant difference in favor of one of the methods in pairs ECO-ALG vs AMG-ECO and STRONG vs AMG. However, we note that in both Benchmarks I and II several graphs exhibited that AMG versions yield to the respective matching for large $k$. The problem is eliminated when we stabilize $\rho$ by using more relaxations according to Theorem 4.2 in \cite{safro:relaxml}.
We cannot present here the exact comparison of coarsening running times because their underlying implementations are very different. Theoretically, however, if in both matching and AMG configurations the algebraic distance is used and when the order of interpolation in AMG is limited by 2 (and usually it is 1, meaning that the coarse graphs are not dense like in 
\cite{ChevalierS09}), the exact complexity of AMG coarsening is not supposed to be bigger than that of matching. 
\section{Conclusions}
We introduced a new coarsening scheme for multilevel graph partitioning based on the AMG coarsening.
One of its most important components, namely, the algebraic distance connectivity measure, has been incorporated into the matching coarsening schemes. 
Both coarsening schemes have been compared under fast and strong configurations of refinement. In addition to known benchmarks, we introduced new potentially hard graphs for large-scale graph partitioning solvers (available at \cite{dimacs10}).
As the main conclusion of this work, we emphasize the success of the proposed AMG coarsening and the algebraic distance connectivity measure between nodes demonstrated on highly irregular instances. One has to take into account the trade-off between increased running time when using algebraic distance and improved quality of the partitions. The increasing running time becomes less tangible with growth of graph size compared with the complexity of the refinement phase.  

Many opportunities remain to improve the coarsening schemes for graph partitioning. We demonstrated the crucial importance of the connectivity strength metrics (especially for fast versions of the algorithm) which raises the question how one can use these metrics at the uncoarsening phase. Preliminary experiments show that this has the potential to improve fast versions even more. 
Another issue that requires more insight is related to the balancing of AMG aggregates. We observed a number of examples for which the unbalanced coarsening produces noticeably better results.

\vfill
\small
\bibliographystyle{plain}
\bibliography{mlpart,quellen}

\begin{thebibliography}{10}

\bibitem{dimacs10}
David~A. Bader, Henning Meyerhenke, Peter Sanders, and Dorothea Wagner.
\newblock 10th {D}{I}{M}{A}{C}{S} {I}mplementation {C}hallenge - {G}raph
  {P}artitioning and {G}raph {C}lustering.
\newblock \url{http://www.cc.gatech.edu/dimacs10/}.

\bibitem{BartelGKM10}
Gereon Bartel, Carsten Gutwenger, Karsten Klein, and Petra Mutzel.
\newblock An experimental evaluation of multilevel layout methods.
\newblock In Ulrik Brandes and Sabine Cornelsen, editors, {\em Graph Drawing},
  volume 6502 of {\em Lecture Notes in Computer Science}, pages 80--91.
  Springer, 2010.

\bibitem{bamg-review}
A.~Brandt.
\newblock Multiscale scientific computation: Review 2001.
\newblock In T.~Barth, R.~Haimes, and T.~Chan, editors, {\em Multiscale and
  Multiresolution methods (Proceeding of the Yosemite Educational Symposium,
  October 2000)}. Springer-Verlag, 2001.

\bibitem{buiMoon96}
T.~N. Bui and B.~R. Moon.
\newblock Genetic algorithm and graph partitioning.
\newblock {\em IEEE Trans. Comput.}, 45(7):841--855, 1996.

\bibitem{journals/ipl/BuiJ92}
Thang~Nguyen Bui and Curt Jones.
\newblock Finding good approximate vertex and edge partitions is {N}{P}-hard.
\newblock {\em Inf. Process. Lett.}, 42(3):153--159, 1992.

\bibitem{chen-safro-algdist-full}
Jie Chen and Ilya Safro.
\newblock Algebraic distance on graphs.
\newblock {\em SIAM Journal on Scientific Computing}, 33(6):3468--3490, 2011.

\bibitem{ChevalierS09}
C{\'e}dric Chevalier and Ilya Safro.
\newblock Comparison of coarsening schemes for multilevel graph partitioning.
\newblock In Thomas St{\"u}tzle, editor, {\em LION}, volume 5851 of {\em
  Lecture Notes in Computer Science}, pages 191--205. Springer, 2009.

\bibitem{Dhillon05afast}
Inderjit Dhillon.
\newblock A fast kernel-based multilevel algorithm for graph clustering.
\newblock In {\em In Proceedings of the 11th ACM SIGKDD International
  Conference on Knowledge Discovery and Data Mining}, pages 629--634, 2005.

\bibitem{DH03a}
D.~Drake and S.~Hougardy.
\newblock A simple approximation algorithm for the weighted matching problem.
\newblock {\em Information Processing Letters}, 85:211--213, 2003.

\bibitem{fiduccia1982lth}
C.~M. Fiduccia and R.~M. Mattheyses.
\newblock {A Linear-Time Heuristic for Improving Network Partitions}.
\newblock In {\em 19th Conference on Design Automation}, pages 175--181, 1982.

\bibitem{fjallstrom1998agp}
P.O. Fjallstrom.
\newblock {Algorithms for graph partitioning: A survey}.
\newblock {\em Linkoping Electronic Articles in Computer and Information
  Science}, 3(10), 1998.

\bibitem{kappa}
M.~Holtgrewe, P.~Sanders, and C.~Schulz.
\newblock {Engineering a Scalable High Quality Graph Partitioner}.
\newblock {\em 24th IEEE International Parallal and Distributed Processing
  Symposium}, 2010.

\bibitem{Hu:wavefront}
Y.~F. Hu and J.~A. Scott.
\newblock A multilevel algorithm for wavefront reduction.
\newblock {\em SIAM J. Sci. Comput.}, 23:1352--1375, April 2001.

\bibitem{KarypisK95}
George Karypis and Vipin Kumar.
\newblock Analysis of multilevel graph partitioning.
\newblock In {\em SC}, 1995.

\bibitem{snap}
J.~Lescovec.
\newblock Stanford {N}etwork {A}nalysis {P}ackage ({S}{N}{A}{P}).
\newblock \url{http://snap.stanford.edu/index.html}.

\bibitem{MauSan07}
J.~Maue and P.~Sanders.
\newblock Engineering algorithms for approximate weighted matching.
\newblock In {\em 6th Workshop on Exp. Algorithms ({WEA})}, volume 4525 of {\em
  LNCS}, pages 242--255. Springer, 2007.

\bibitem{meyerhenke2008ndb}
H.~Meyerhenke, B.~Monien, and T.~Sauerwald.
\newblock {A new diffusion-based multilevel algorithm for computing graph
  partitions of very high quality}.
\newblock In {\em IEEE International Symposium on Parallel and Distributed
  Processing, 2008. IPDPS 2008.}, pages 1--13, 2008.

\bibitem{Scotch}
F.~Pellegrini.
\newblock Scotch home page.
\newblock {\url{http://www. labri.fr/pelegrin/scotch}}.

\bibitem{pothen-part}
A.~Pothen, H.~D. Simon, and K.-P. Liou.
\newblock Partitioning sparse matrices with eigenvectors of graphs.
\newblock {\em SIAM J. Matrix Anal. Appl.}, 11(3):430--452, 1990.

\bibitem{doritpart}
D.~Ron, S.~Wishko-Stern, and A.~Brandt.
\newblock An algebraic multigrid based algorithm for bisectioning general
  graphs.
\newblock Technical Report MCS05-01, Department of Computer Science and Applied
  Mathematics, The Weizmann Institute of Science, 2005.

\bibitem{safro:relaxml}
Dorit Ron, Ilya Safro, and Achi Brandt.
\newblock Relaxation-based coarsening and multiscale graph organization.
\newblock {\em Multiscale Modeling {\&} Simulation}, 9(1):407--423, 2011.

\bibitem{safro2005}
I.~Safro, D.~Ron, and A.~Brandt.
\newblock Multilevel algorithms for linear ordering problems.
\newblock {\em Journal of Experimental Algorithmics}, 13:1.4--1.20, 2008.

\bibitem{hardpart-site}
I.~Safro, P.~Sanders, and C.~Schulz.
\newblock Benchmark with {P}otentially {H}ard {G}raphs for {P}artitioning
  {P}roblem.
\newblock \url{http://www.mcs.anl.gov/~safro/hardpart.html}.

\bibitem{kaffpaE}
P.~Sanders and C.~Schulz.
\newblock {Distributed Evolutionary Graph Partitioning}.
\newblock {\em 12th Workshop on Algorithm Engineering and Experimentation},
  2011.

\bibitem{kaffpa}
P.~Sanders and C.~Schulz.
\newblock {Engineering Multilevel Graph Partitioning Algorithms}.
\newblock {\em 19th European Symposium on Algorithms}, 2011.

\bibitem{SchKarKum00}
K.~Schloegel, G.~Karypis, and V.~Kumar.
\newblock Graph partitioning for high performance scientific simulations.
\newblock In J.~Dongarra and et~al., editors, {\em CRPC Par. Comp. Handbook}.
  Morgan Kaufmann, 2000.

\bibitem{sharon}
E.~Sharon, A.~Brandt, and R.~Basri.
\newblock Fast multiscale image segmentation.
\newblock In {\em Proceedings IEEE Conference on Computer Vision and Pattern
  Recognition}, pages 70--77, 2000.

\bibitem{soper2004combined}
A.J. Soper, C.~Walshaw, and M.~Cross.
\newblock A combined evolutionary search and multilevel optimisation approach
  to graph-partitioning.
\newblock {\em Journal of Global Optimization}, 29(2):225--241, 2004.

\bibitem{stock2006strategic}
L.E. Stock.
\newblock {\em Strategic Logistics Management}.
\newblock Cram101 Textbook Outlines. Lightning Source Inc, 2006.

\bibitem{multimode}
Lei Tang, Huan Liu, Jianping Zhang, and Zohreh Nazeri.
\newblock Community evolution in dynamic multi-mode networks.
\newblock In {\em KDD}, pages 677--685, 2008.

\bibitem{mgbooktrott}
Ulrich Trottenberg and Anton Schuller.
\newblock {\em Multigrid}.
\newblock Academic Press, Inc., Orlando, FL, USA, 2001.

\bibitem{walshaw2004multilevel}
C.~Walshaw.
\newblock {Multilevel refinement for combinatorial optimisation problems}.
\newblock {\em Annals of Operations Research}, 131(1):325--372, 2004.

\bibitem{Walshaw07}
C.~Walshaw and M.~Cross.
\newblock {JOSTLE: Parallel Multilevel Graph-Partitioning Software -- An
  Overview}.
\newblock In F.~Magoules, editor, {\em {Mesh Partitioning Techniques and Domain
  Decomposition Techniques}}, pages 27--58. Civil-Comp Ltd., 2007.
\newblock (Invited chapter).

\end{thebibliography}
\vspace*{1cm}
\hspace*{1.5in}{\scriptsize\framebox{\parbox{2.4in}{
The submitted manuscript has been created in part by UChicago Argonne, LLC, Operator of Argonne National Laboratory (``Argonne'').  Argonne, a U.S. Department of Energy Office of Science laboratory, is operated under Contract No. DE-AC02-06CH11357.  The U.S. Government retains for itself, and others acting on its behalf, a paid-up nonexclusive, irrevocable worldwide license in said article to reproduce, prepare derivative works, distribute copies to the public, and perform publicly and display publicly, by or on behalf of the Government.
}}}
\end{document}